\documentclass[twocolumn]{aastex62}
\usepackage[outdir=./]{epstopdf}
\received{}
\revised{}
\accepted{}

\submitjournal{ApJ}

\shorttitle{Survey for SySt in the Magellanic Clouds - first discoveries}
\shortauthors{I\l{}kiewicz al.}

\begin{document}

\title{A deep survey for symbiotic stars in the Magellanic Clouds - 1. Methodology and first discoveries in the SMC}

\correspondingauthor{Krystian I\l{}kiewicz}
\email{ilkiewicz@camk.edu.pl}

\author{Krystian I\l{}kiewicz}
\affil{N. Copernicus Astronomical Center, Polish Academy of Sciences \\
Bartycka 18, PL~00--716 Warsaw, Poland}

\author{Joanna Miko\l{}ajewska}
\affil{N. Copernicus Astronomical Center, Polish Academy of Sciences \\
Bartycka 18, PL~00--716 Warsaw, Poland}

\author{Michael M. Shara}
\affil{Department of Astrophysics, American Museum of Natural History \\
CPW \& 79$^\mathrm{th}$ street, New York, NY 10024, USA}
\affil{Institute of Astronomy, The Observatories  \\
Madingley Road, Cambridge CB3 0HA, UK}

\author{Andrzej Udalski}
\affiliation{Warsaw University Observatory \\
Al. Ujazdowskie 4, 00-478 Warsaw, Poland}

\author{Katarzyna Drozd}
\affil{N. Copernicus Astronomical Center, Polish Academy of Sciences \\
Bartycka 18, PL~00--716 Warsaw, Poland}

\author{Jacqueline K. Faherty}
\affil{Department of Astrophysics, American Museum of Natural History \\
CPW \& 79$^\mathrm{th}$ street, New York, NY 10024, USA}

\begin{abstract}

We have initiated a survey aimed at locating  a nearly complete sample of classical symbiotic stars (SySt) in the Magellanic Clouds. Such a sample is nearly impossible to obtain in the Milky Way, and is essential to constrain the formation, evolution and demise of these strongly interacting, evolved binary stars. We have imaged both Clouds in H$\alpha$ and \mbox{He\,{\sc ii}\,4686} narrow-band filters deeply enough to detect all known symbiotic stars. While \mbox{He\,{\sc ii}\,4686} is not present in all SySt, our method should yield a high success rate because the mimics of SySt are not as likely as true symbiotics to show this emission line. We demonstrate the viability of our method through the discovery and characterization of three new SySt in the Small Magellanic Cloud: 2MASS J00411657-7233253, 2MASS J01104404-7208464 and 2MASS J01113745-7159023. Enigmatic variability was observed in 2MASS J01113745-7159023, where changes in the amplitude of its quasi-periodic variability may suggest an enhanced mass transfer rate during a periastron passage on an elliptical orbit. 2MASS J01104404-7208464 is an ellipsoidal variable with an orbital period of 403\,d.

\end{abstract}


\section{Introduction} \label{sec:intro}

Symbiotic stars (SySt) are the longest orbital period interacting binaries composed of an evolved cool giant (either a normal red giant (RG) in S-type, or a Mira surrounded by an opaque dust shell in D-type SySt) and an accreting hot, and luminous companion (usually a white dwarf, WD). The interacting stars are surrounded by rich and complex circumstellar environments, including both ionized and neutral regions, dust forming regions, accretion/excretion disks, interacting winds, bipolar outflows and jets. SySt are very important and luminous tracers of the late phases of low- and medium-mass binary star evolution, and excellent laboratories to test models of close binary evolution. They are also promising nurseries for type Ia supernovae, regardless of whether the path to the thermonuclear explosion of a Chandrasekhar mass CO WD is through the accretion (single degenerate, SD) scenario or through merging a double WD system (double degenerate, DD) scenario. The most recent extensive review of SySt is in \citet{2012BaltA..21....5M}. 

While about 300 Galactic SySt are known (e.g. \citealt{2000A&AS..146..407B}; \citealt{2013MNRAS.432.3186M}; \citealt{2014MNRAS.440.1410M}; \citealt{2014A&A...567A..49R}, and references therein), and a few dozen are relatively well studied, their distances (and hence their component luminosities and other distance-related parameters) remain poorly determined. This makes confrontation of theoretical models of SySt with observed parameters of real SySt, to test theories of their evolution and interactions, very challenging. A much more complete and systematic search for SySt in multiple galaxies of different types is essential to provide deep and complete samples suitable for tests of binary evolution theory. 

Fortunately, ten bright SySt have been detected in each of the Magellanic Clouds (\citealt{2000A&AS..146..407B}; \citealt{2014MNRAS.444L..11M}; \citealt{2018MNRAS.476.2605I} and references therein), and recently SySt have became detectable even at several hundred kpc in Local Group Galaxies (\citealt{2008MNRAS.391L..84G}; \citealt{2009MNRAS.395.1121K}; \citealt{2014MNRAS.444..586M}; \citealt{2017MNRAS.465.1699M}; \citealt{2017A&A...606A.110I}). 

Almost all SySt discovered before 2000 were found serendipitously. Since then, increasing numbers were located by using H$\alpha$ surveys of the Galaxy (\citealt{2014MNRAS.440.1410M}; \citealt{2014A&A...567A..49R}, and references therein) and the Local Group of galaxies (\citealt{2015EAS....71..199M}; \citealt{2017MNRAS.465.1699M}). Success rates, defined as the fraction of spectrographically observed candidates to confirmed SySt in these surveys, are typically 10\% or less. For this reason the survey we have initiated is based on both narrowband H$\alpha$ and \mbox{He\,{\sc ii}\,4686} imaging. While not all SySt display \mbox{He\,{\sc ii}\,4686} emission, a major fraction do, e.g.,  68\% of SySt included in the spectrophotometric atlas of symbiotic stars published by \citet{2002A&A...383..188M}  have \mbox{He\,{\sc ii}\,4686}/H$\beta$ $>~0.1$, and this will greatly reduce the waste of spectrographic time spent on mimics. 

In this paper we present the methodology of our survey, and the first results of our search of the Small Magellanic Cloud (SMC) aiming at finding new SySt. Section 2 describes the survey and our candidate selection method. In Section 3 we present our spectroscopic observations. Characterization of the newly discovered SySt is presented in Section 4. Our results are summarized in section 5.

\section{Candidate selection}

We carried out photometric observations with the Swope 1.0m telescope and E2V~CCD231-84 CCD camera on the nights of 2016.09.07-2016.09.22. The survey consisted of observations in H$\alpha$ and \mbox{He\,{\sc ii}\,4686} narrow-band filters and a broad-band $R$ and $B$ filters. For each field we stacked 5 images with 45 second exposure time in $R$ and $B$ filters and 450 second exposure times in H$\alpha$ and \mbox{He\,{\sc ii}\,4686} filters. This resulted in a limiting magnitude of $\sim$18.5~mag in each filter, corresponding to absolute magnitudes $\sim$0 in the LMC and $\sim$0.3 in the SMC. The unceratinities in the measured magnitudes are $\sim 0.02$~mag in all filters.

The E2V~CCD231-84 CCD camera is build from four separate quadrants. We corrected the images for differences in linearity of these quadrants, and for the shutter pattern which was a result of the finite shutter speed. The images were reduced using the standard IRAF procedures. The astrometric calibration was done using Astrometry.net software \citep{2010AJ....139.1782L}. The images were combined using SWarp \citep{2002ASPC..281..228B}. Point-spread function fitting photometry was done using PSFEx \citep{2011ASPC..442..435B} and SExtractor \citep{1996A&AS..117..393B}. 

\begin{figure*}
\centering
\includegraphics[width=0.3\hsize]{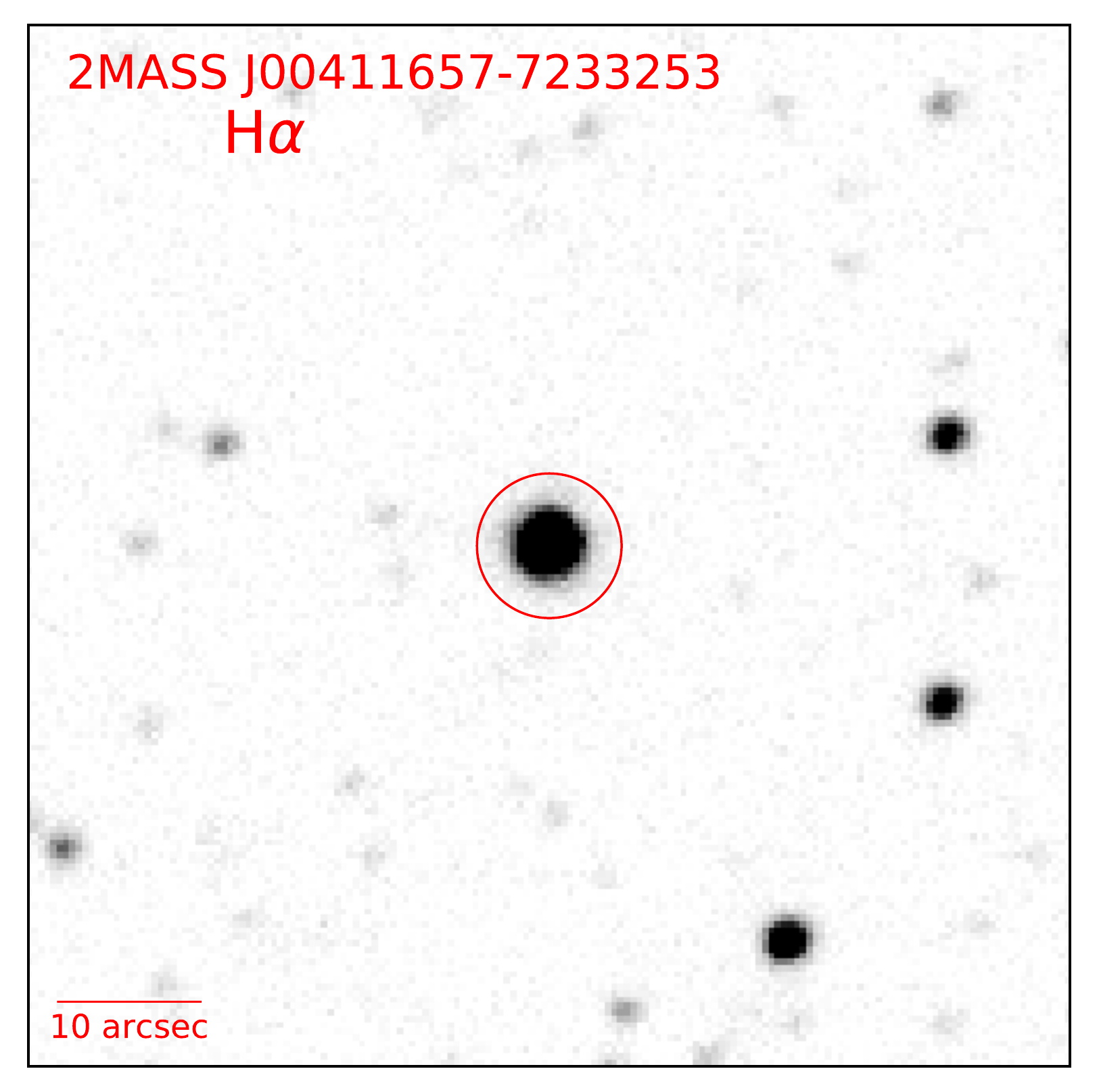}
\includegraphics[width=0.3\hsize]{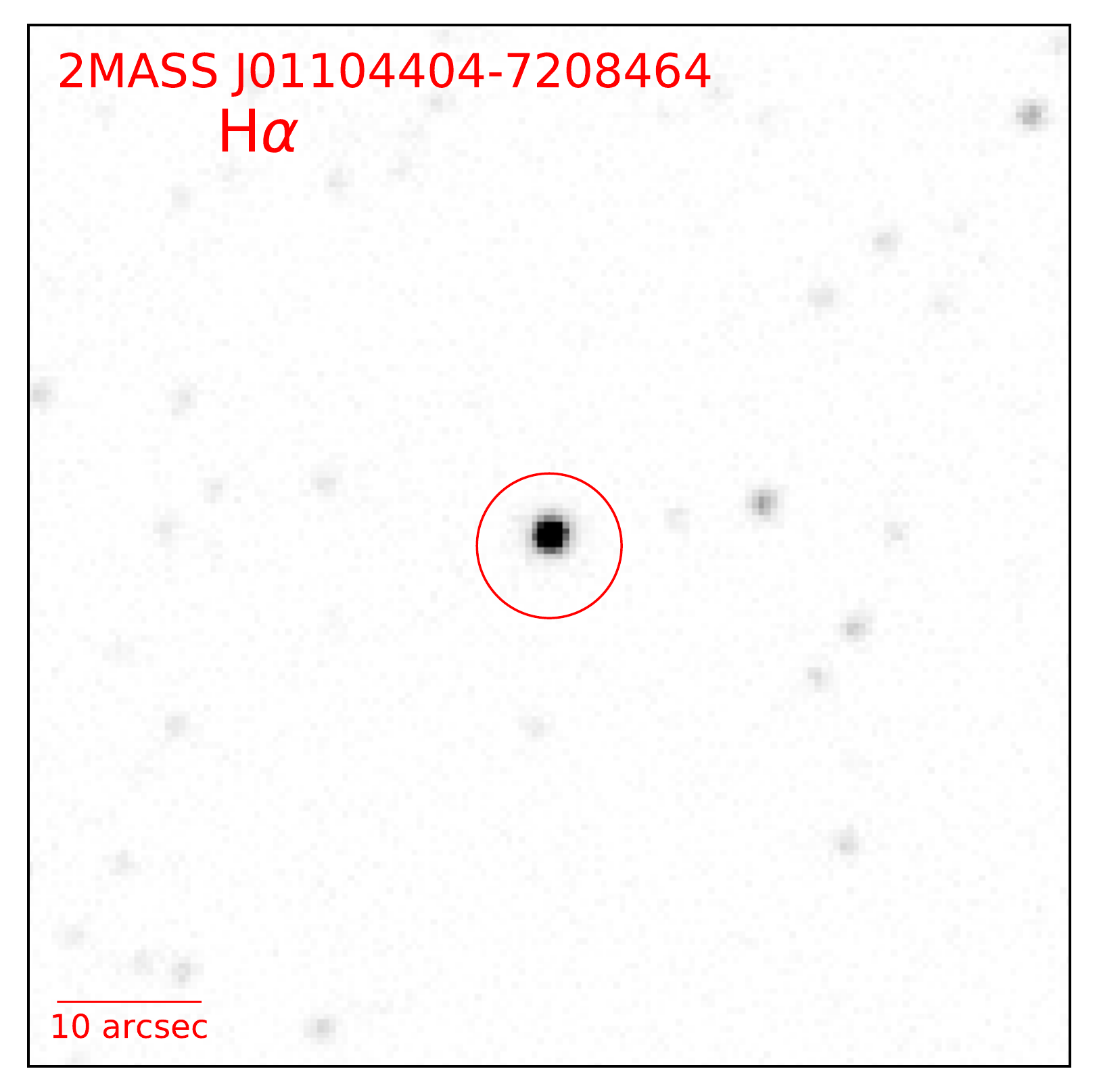}
\includegraphics[width=0.3\hsize]{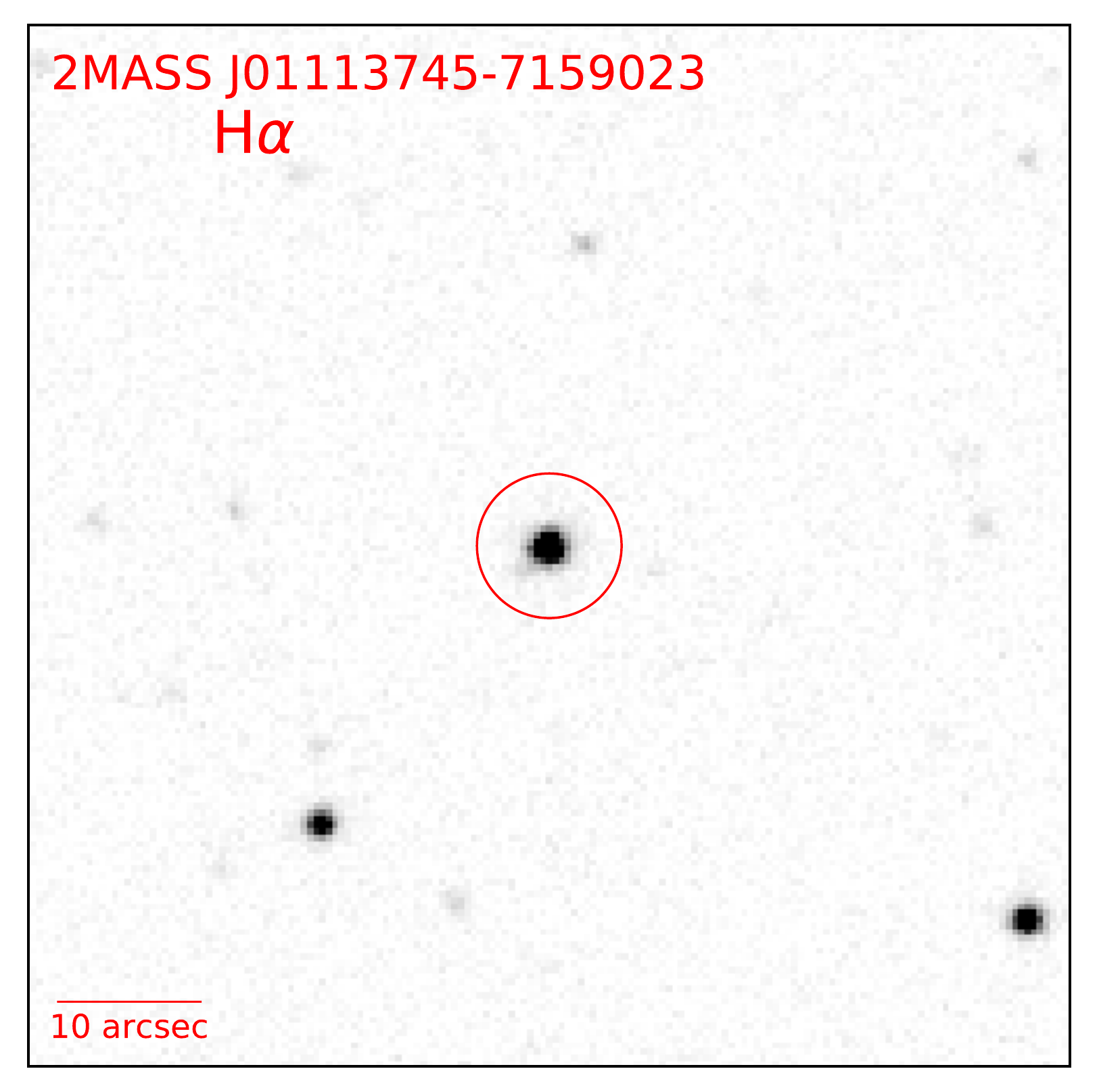}
\includegraphics[width=0.3\hsize]{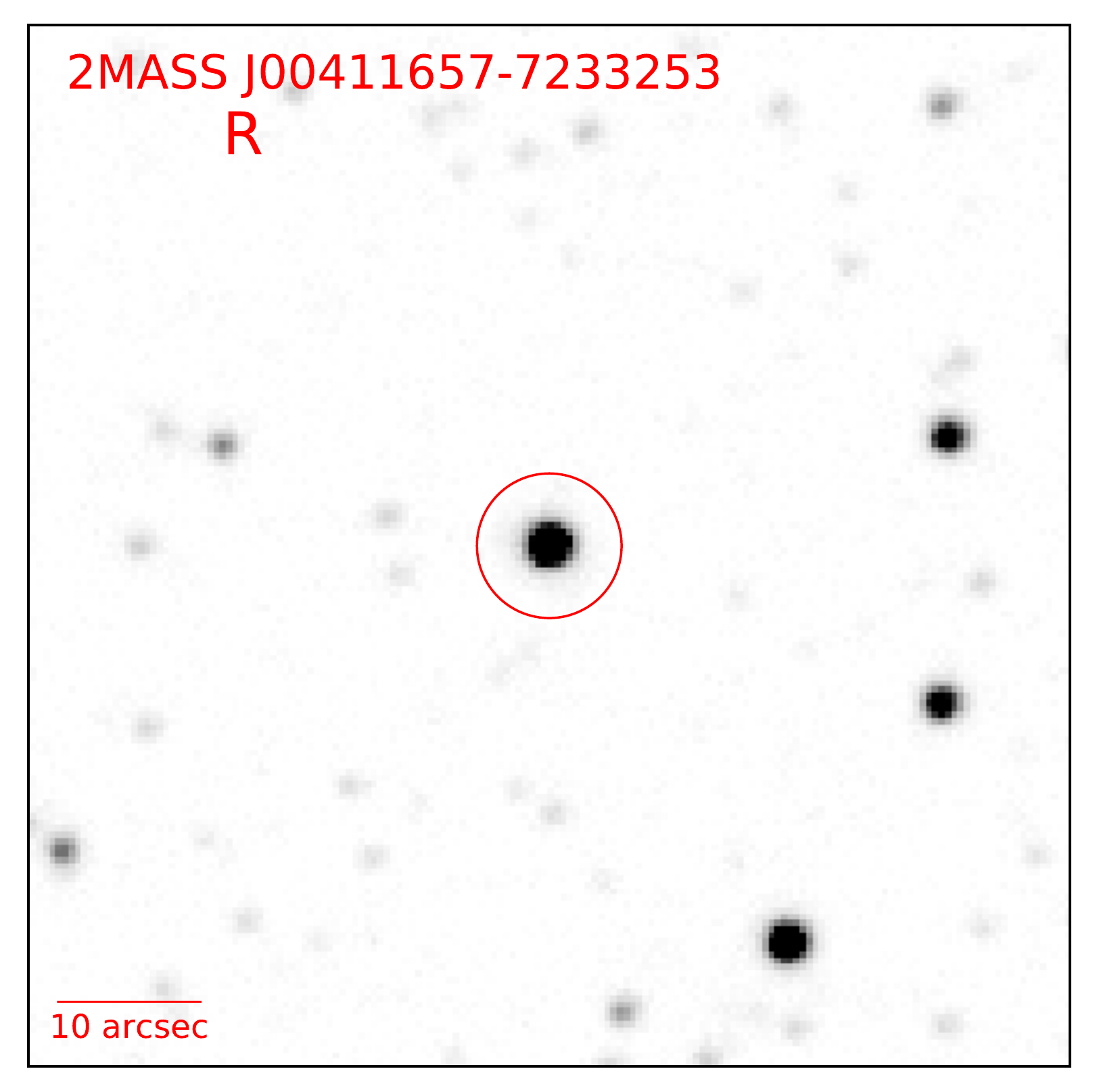}
\includegraphics[width=0.3\hsize]{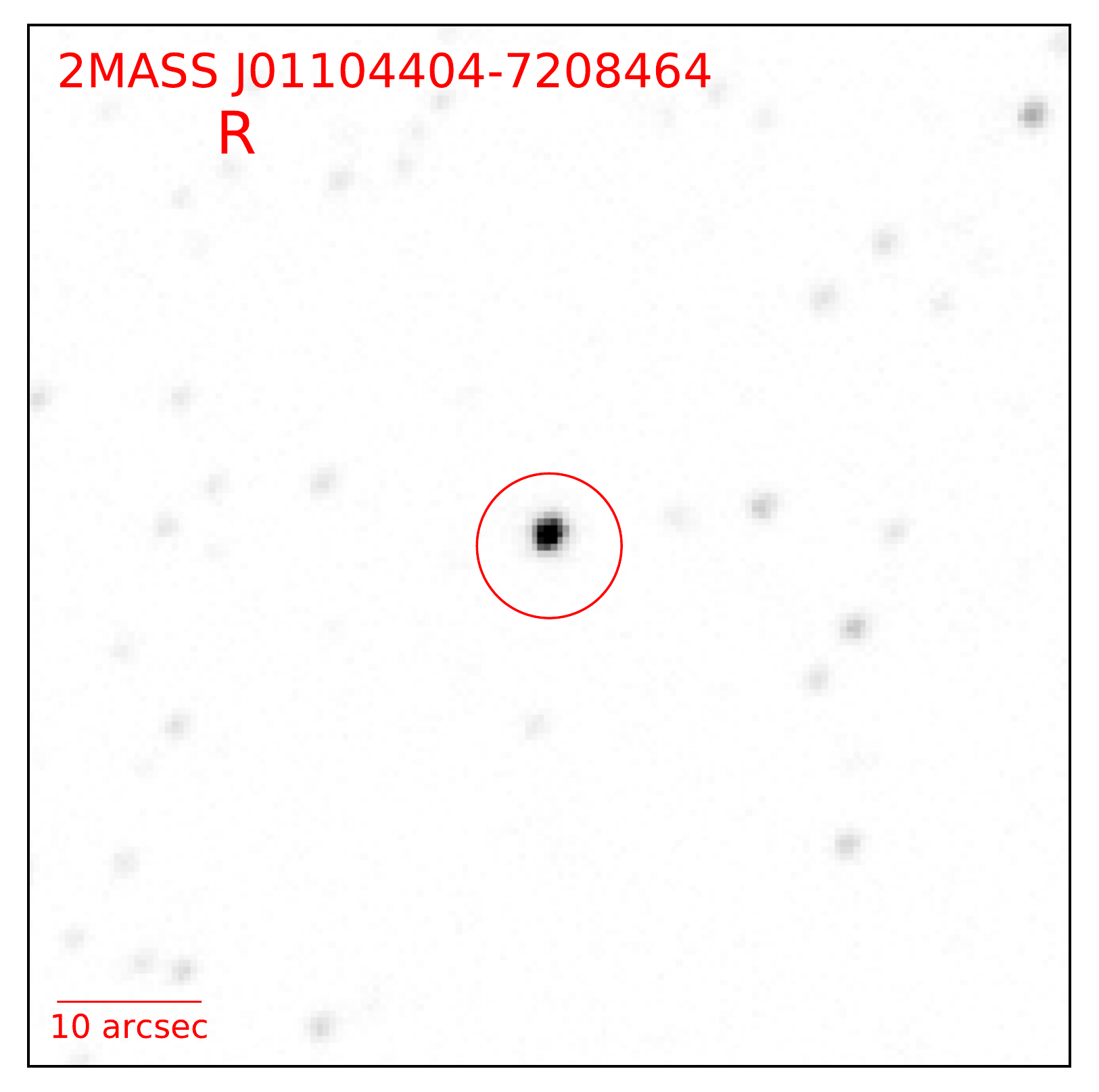} 
\includegraphics[width=0.3\hsize]{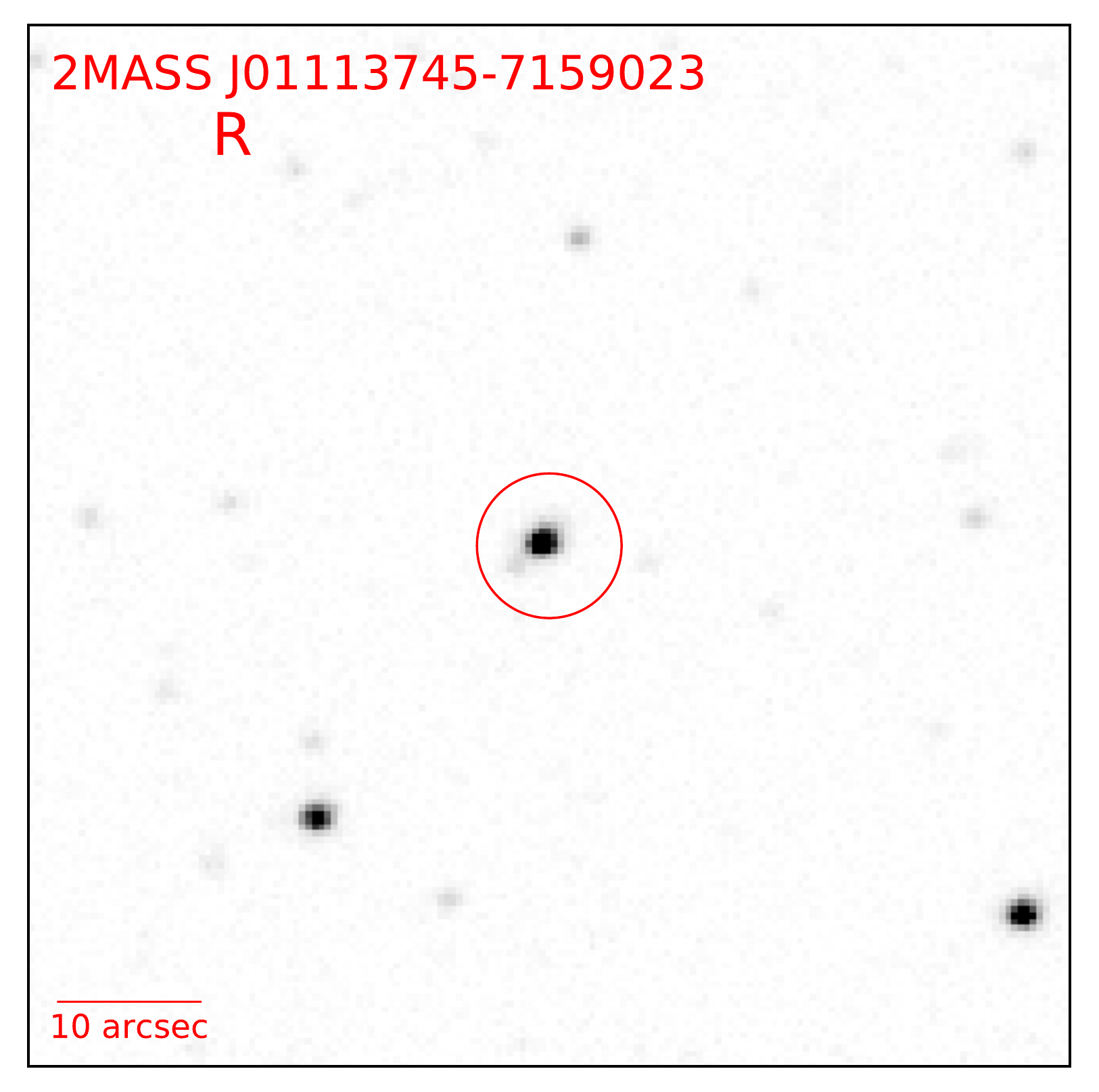}
\includegraphics[width=0.3\hsize]{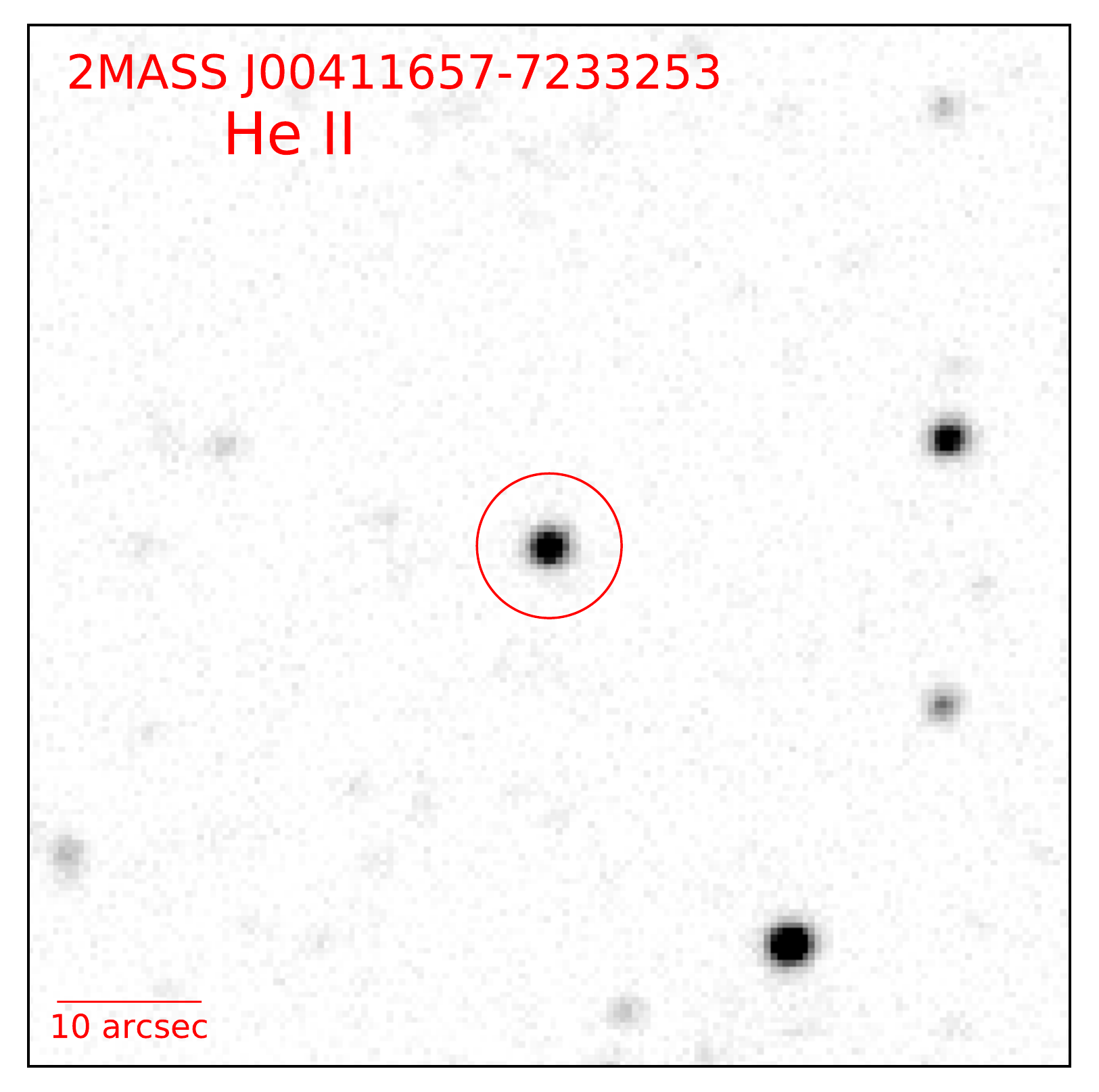}
\includegraphics[width=0.3\hsize]{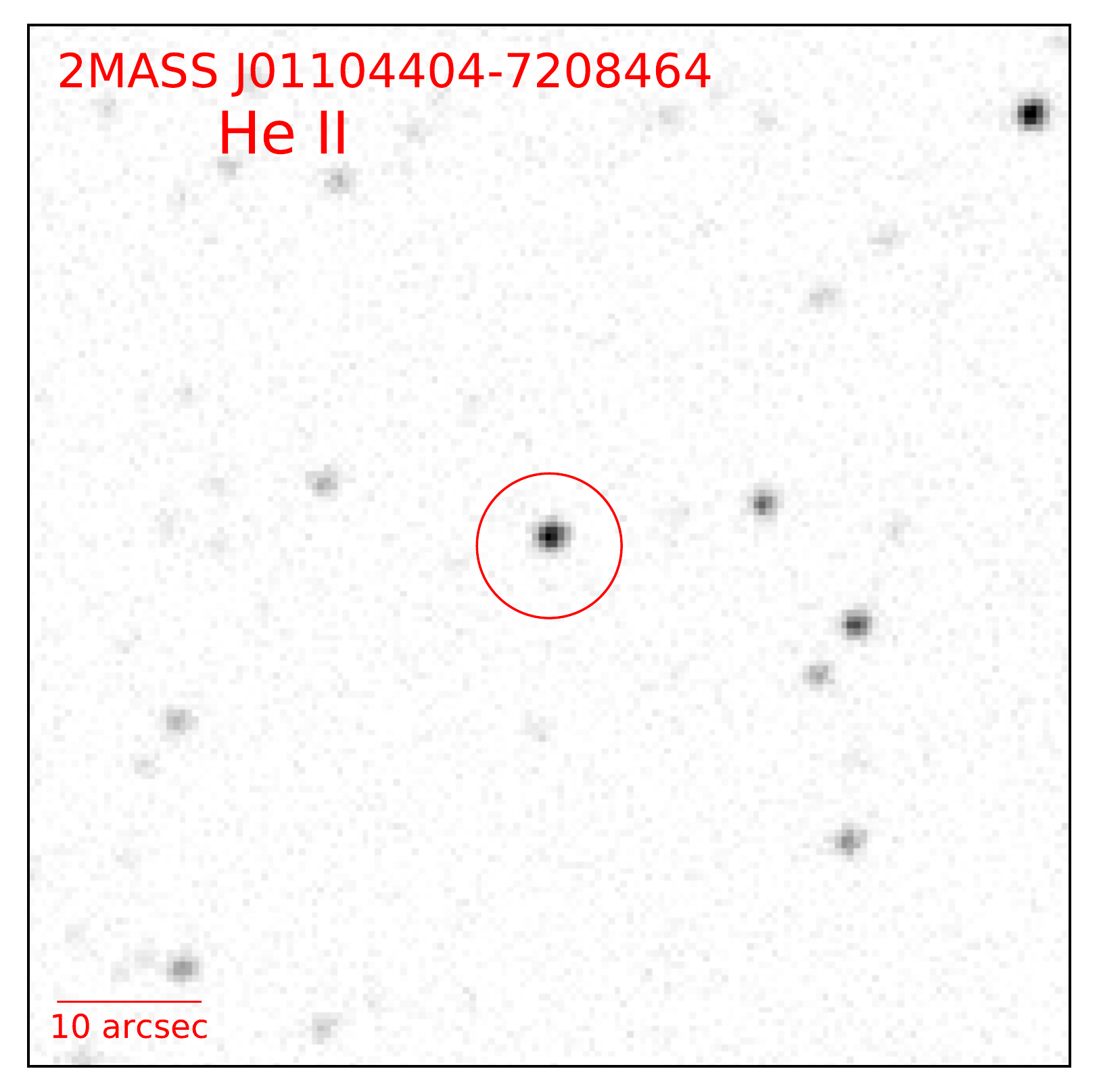}
\includegraphics[width=0.3\hsize]{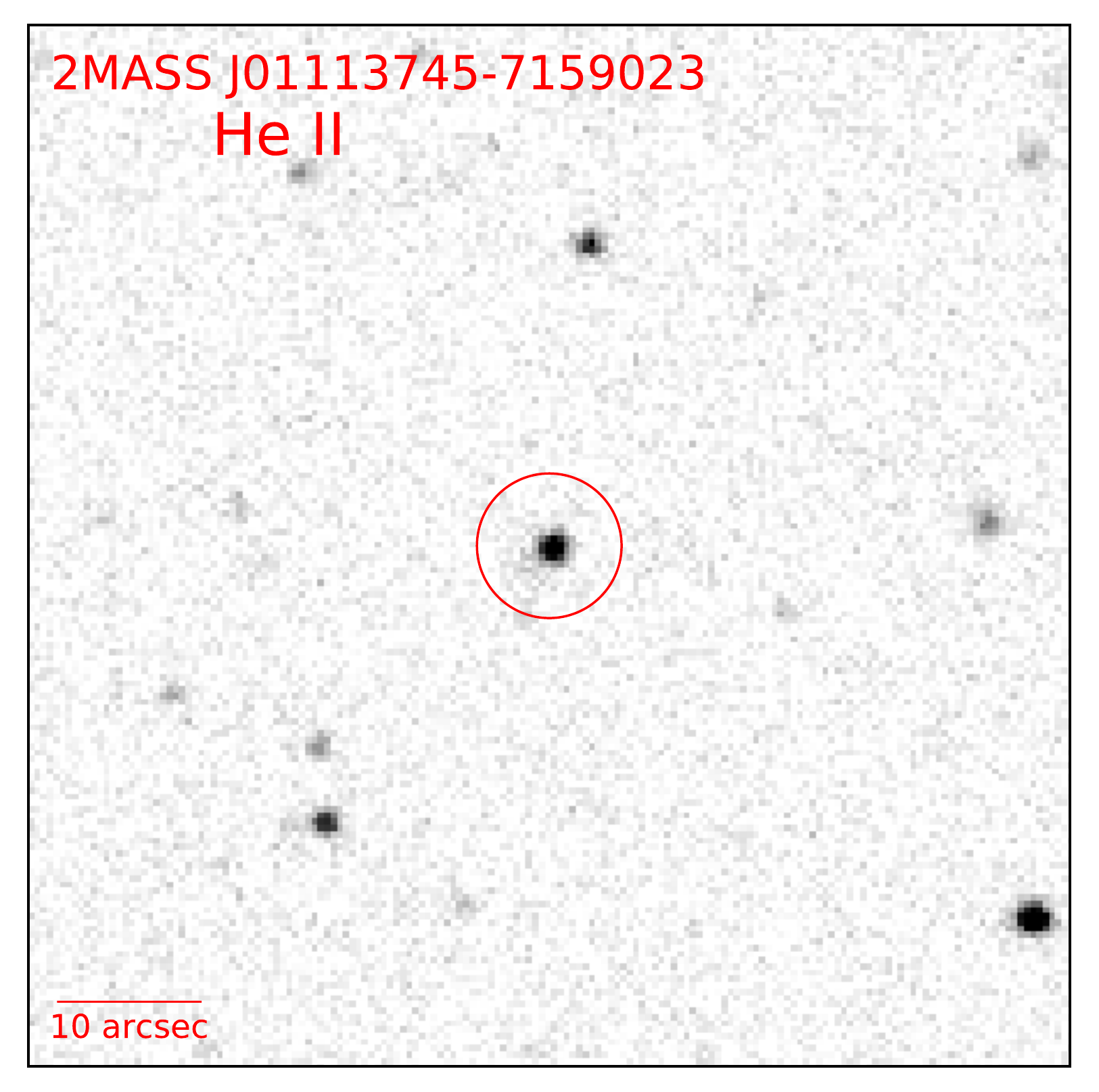}
\includegraphics[width=0.3\hsize]{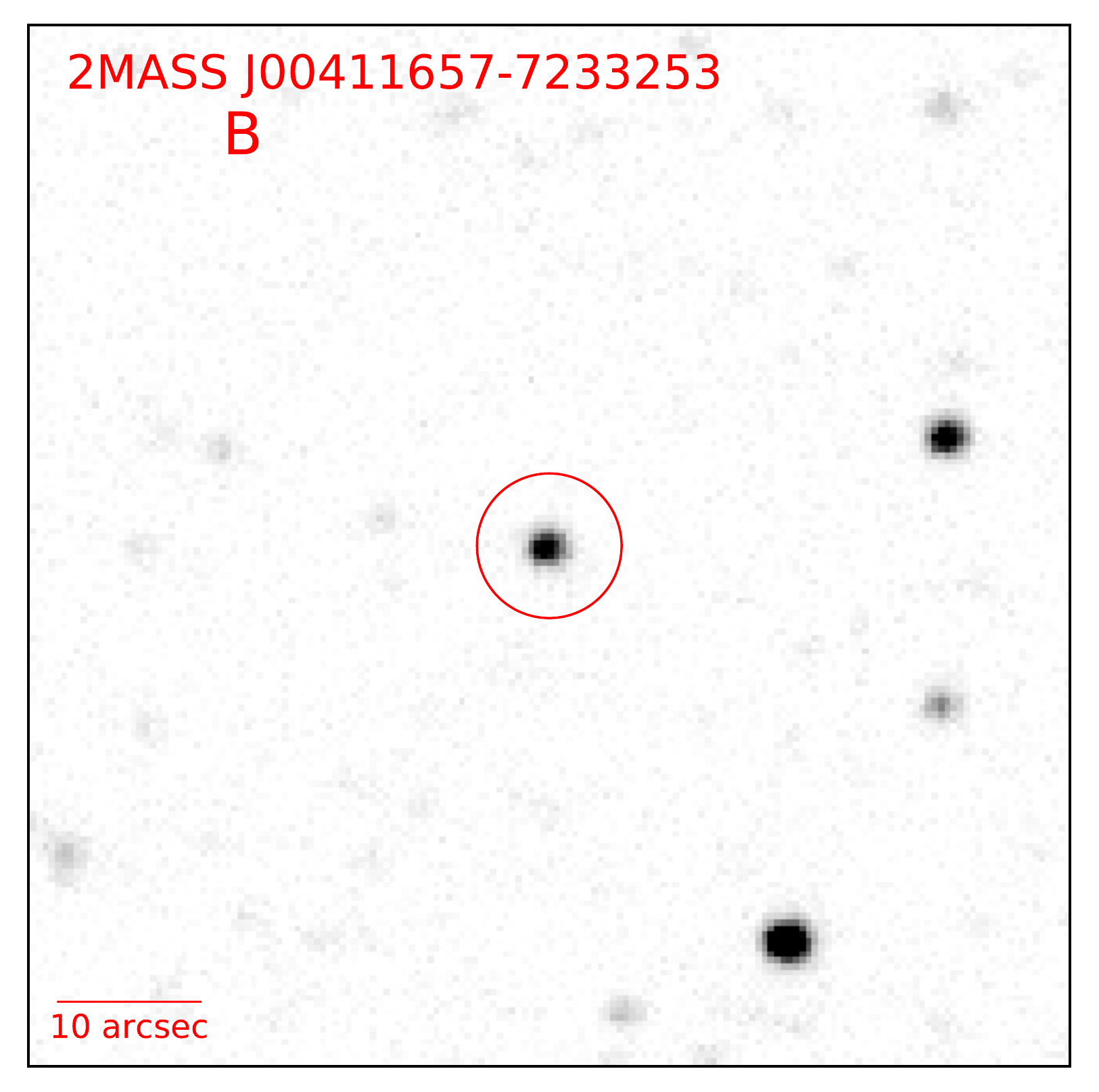}
\includegraphics[width=0.3\hsize]{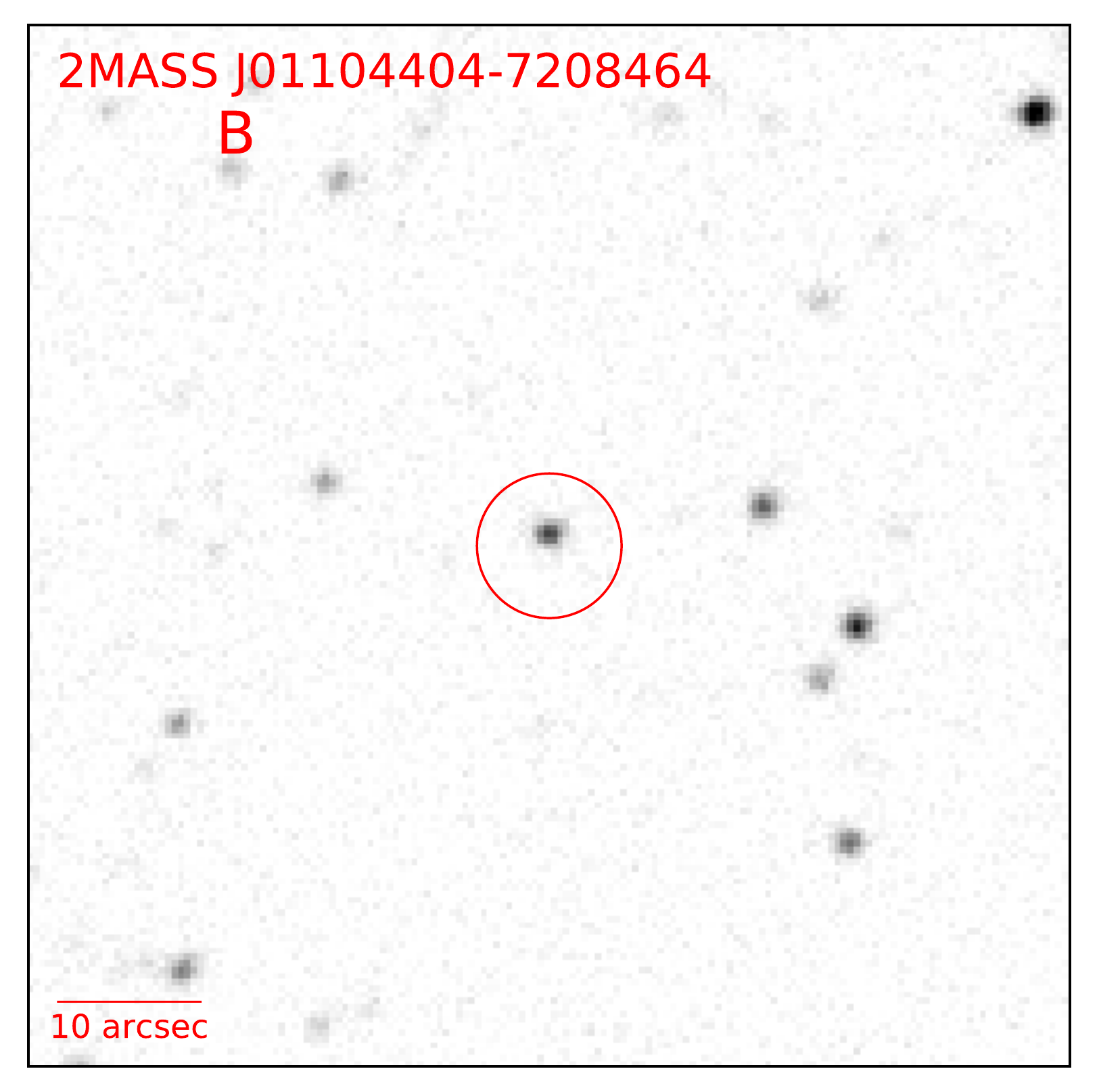}
\includegraphics[width=0.3\hsize]{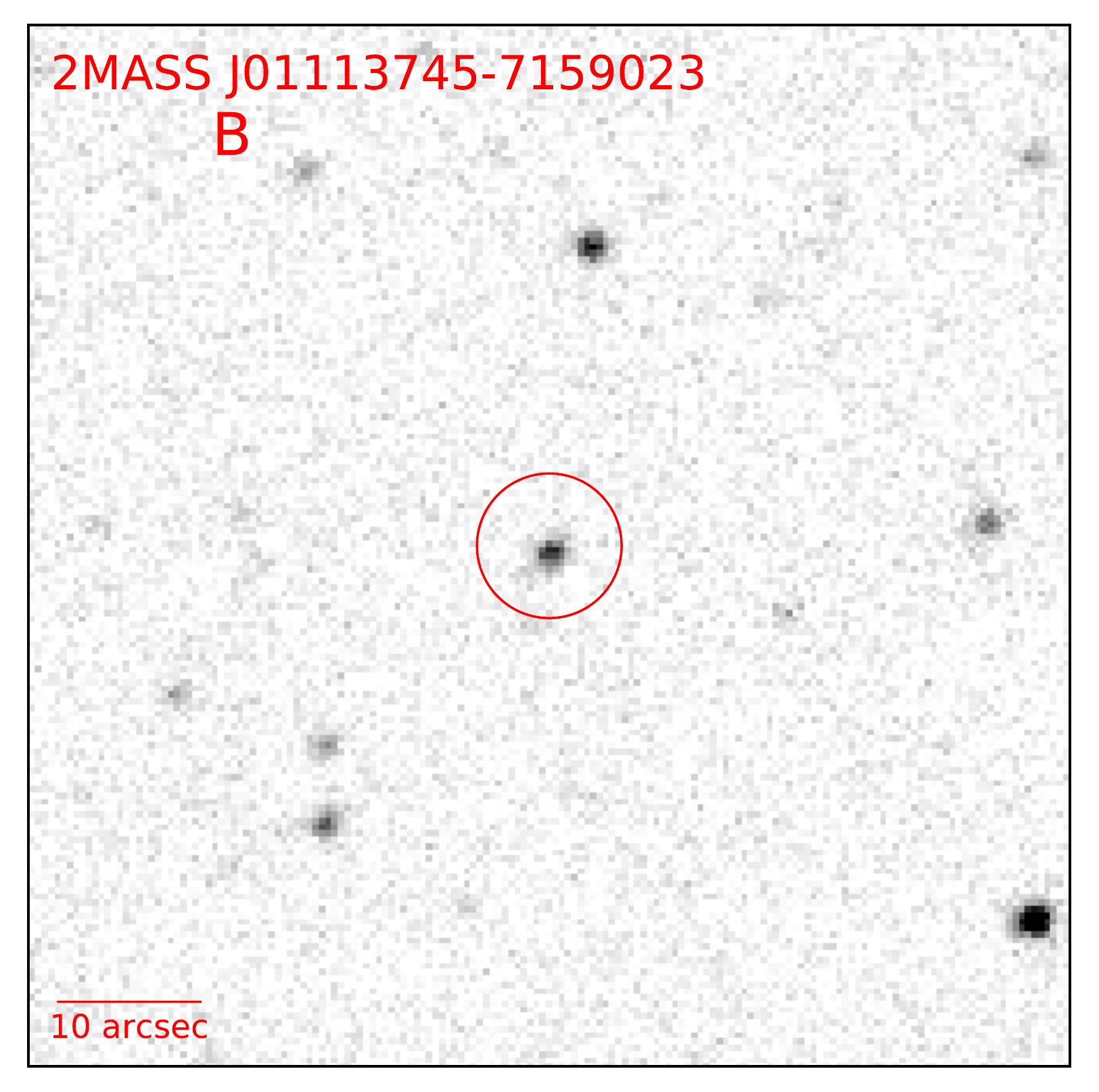}

\caption{Swope images of the new SySt.}
 \label{fig:swope_imgs}
\end{figure*}

We selected the SySt candidates among the stars that show the colors \mbox{He\,{\sc ii}}$-B< -0.1$~mag  and H$\alpha-R<-1.0$~mag. These colors effectively isolate emission line stars with equivalent widths of 10$\AA$ and 100$\AA$, respectively. While not all SySt show the  \mbox{He\,{\sc ii}} line, the stars that mimic SySt are less likely to show this line \citep{2014MNRAS.440.1410M}, which will result in a smaller contamination rate. The Swope images of the first three successful SySt candidates are presented in Fig.~\ref{fig:swope_imgs}.

\section{SALT spectroscopy}

\begin{figure*}[h!]
\includegraphics[width=1.0\hsize]{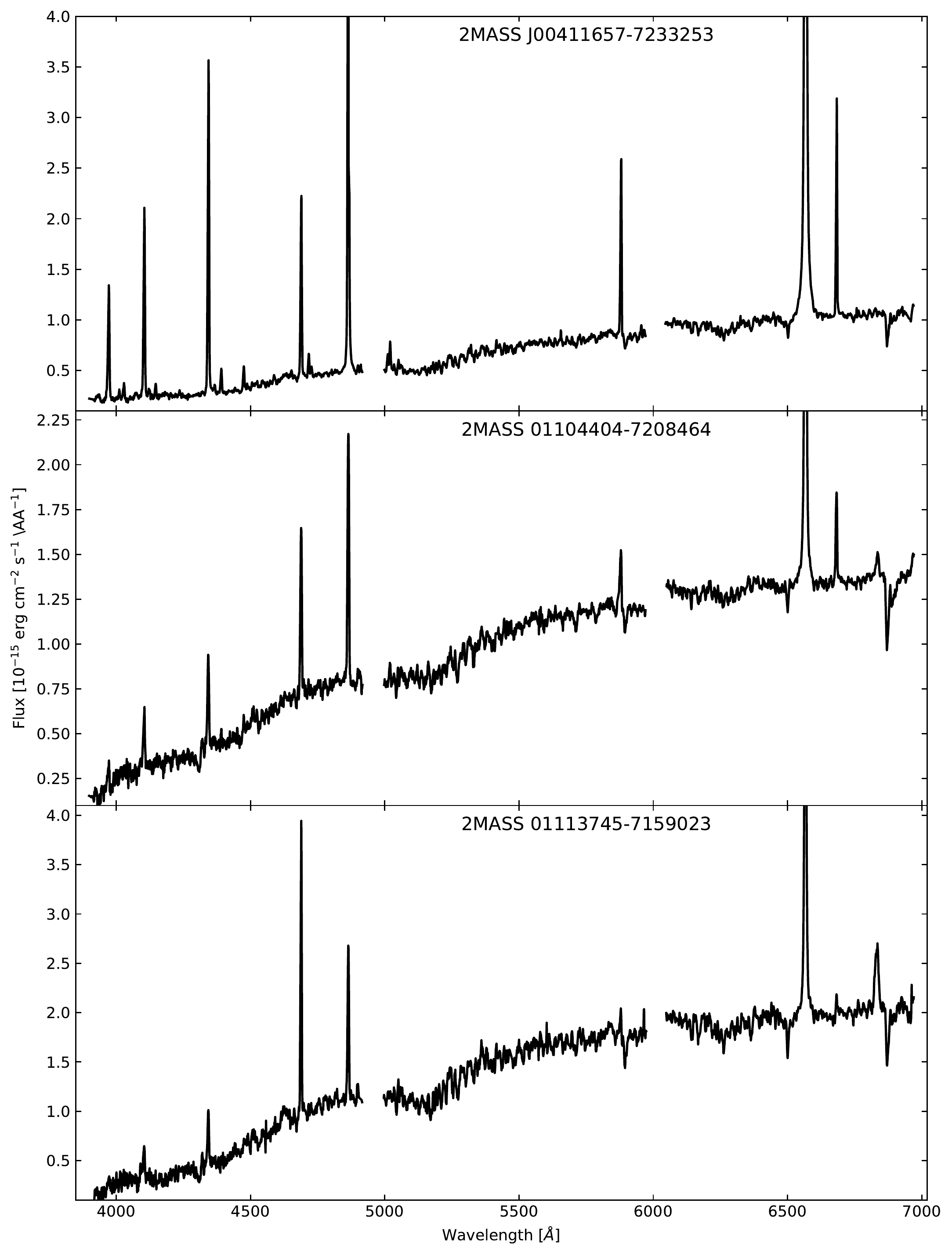}
\caption{SALT spectra of the new SySt.}
\label{fig:salt_spec}
\end{figure*}

We carried out spectroscopic followup of ten uncrowded candidates in the SMC with the  Southern  African  Large  Telescope (SALT; \citealt{2006MNRAS.372..151O}) under  programme  2017-1-SCI-049  (PI:  I\l{}lkiewicz). We used the   Robert-Stobie  Spectrograph  (\citealt{2003SPIE.4841.1463B}; \citealt{2003SPIE.4841.1634K}) with a slit width of 1.5 arcsec and a PG0900 grating, resulting  in  a  resolution  of  R$\simeq$1000.  The wavelength covered in each spectrum was 3900--6900\AA. The reduction  was  carried out  with the standard IRAF tasks and the pysalt package  \citep{2010SPIE.7737E..25C}. The spectra were flux calibrated by scaling them to the known $V$ magnitudes of the candidates. The log of observations is presented in Table~\ref{tab:log}. The spectra of confirmed SySt are presented in Fig.~\ref{fig:salt_spec}. The fluxes of emission lines are presented in Table~\ref{tab:fluxes}. The fluxes have uncertainties of 15\% for the strong lines and 30\% for the weak lines. 

\begin{table*}
\centering
\caption{The log of SALT observations, OGLE $V$ magnitudes used for scaling the spectra and the Swope magnitudes from our survey.} \label{tab:log}
\begin{tabular}{cccccccc}
\tablewidth{0pt}
\hline
\hline
 &		Date &	MJD	&	$V$ [mag]	& $B$ [mag] & \mbox{He\,{\sc ii}} [mag] & $R$ [mag] & H$\alpha$ [mag]	\\
\hline
\decimals
2MASS J00411657-7233253 & 2016-11-11 & 57704 & 16.8 & 16.60 & 16.33 & 15.16 & 13.35 \\
2MASS J01104404-7208464 & 2016-11-11 & 57704 & 16.4 & 17.53 & 17.16 & 15.82 & 15.29 \\
2MASS J01113745-7159023 & 2016-11-12 & 57705 & 16.0 & 17.77 & 16.33 & 15.16 & 13.35 \\
\hline
\end{tabular}
\end{table*}

\begin{table*}
\centering
\caption{Observed emission line fluxes in the new SySt.} \label{tab:fluxes}
\begin{tabular}{cccc}
\tablewidth{0pt}
\hline
\hline
2MASS &	 J00411657-7233253	& J01104404-7208464	& J01113745-7159023	\\
\hline
ID & \multicolumn{3}{c}{Flux [$10^{-15}$~erg~cm$^{-2}$~s$^{-1}$] }\\
\hline
H$_\epsilon$	&	7.2	&	0.9	&		\\
\mbox{He\,{\sc i}} 4026	&	1.1	&		&		\\
H$_\delta$	&	11	&	2.3	&	2.9	\\
\mbox{He\,{\sc i}} 4143	&	0.8	&		&		\\
H$_\gamma$	&	19	&	3.5	&	3.3	\\
\mbox{He\,{\sc i}} 4387	&	1.5	&		&		\\
\mbox{He\,{\sc i}} 4471	&	1.2	&		&		\\
\mbox{He\,{\sc ii}} 4686	&	9.8	&	6.1	&	15	\\
\mbox{He\,{\sc i}} 4713	&	1.2	&		&		\\
H$_\beta$	&	37	&	9.7	&	9.4	\\
\mbox{He\,{\sc i}} 5876	&	8.8	&	2.6	&	1.2	\\
H$_\alpha$	&	163	&	50	&	52	\\
\mbox{He\,{\sc i}} 6678	&	9.4	&	3.1	&	0.9	\\
\mbox{O\,{\sc vi}} 6825	&		&	1.4	&	9.5	\\

\hline
\end{tabular}
\end{table*}

\section{New symbiotic stars}

\begin{figure}
\centering
\includegraphics[width=0.99\hsize]{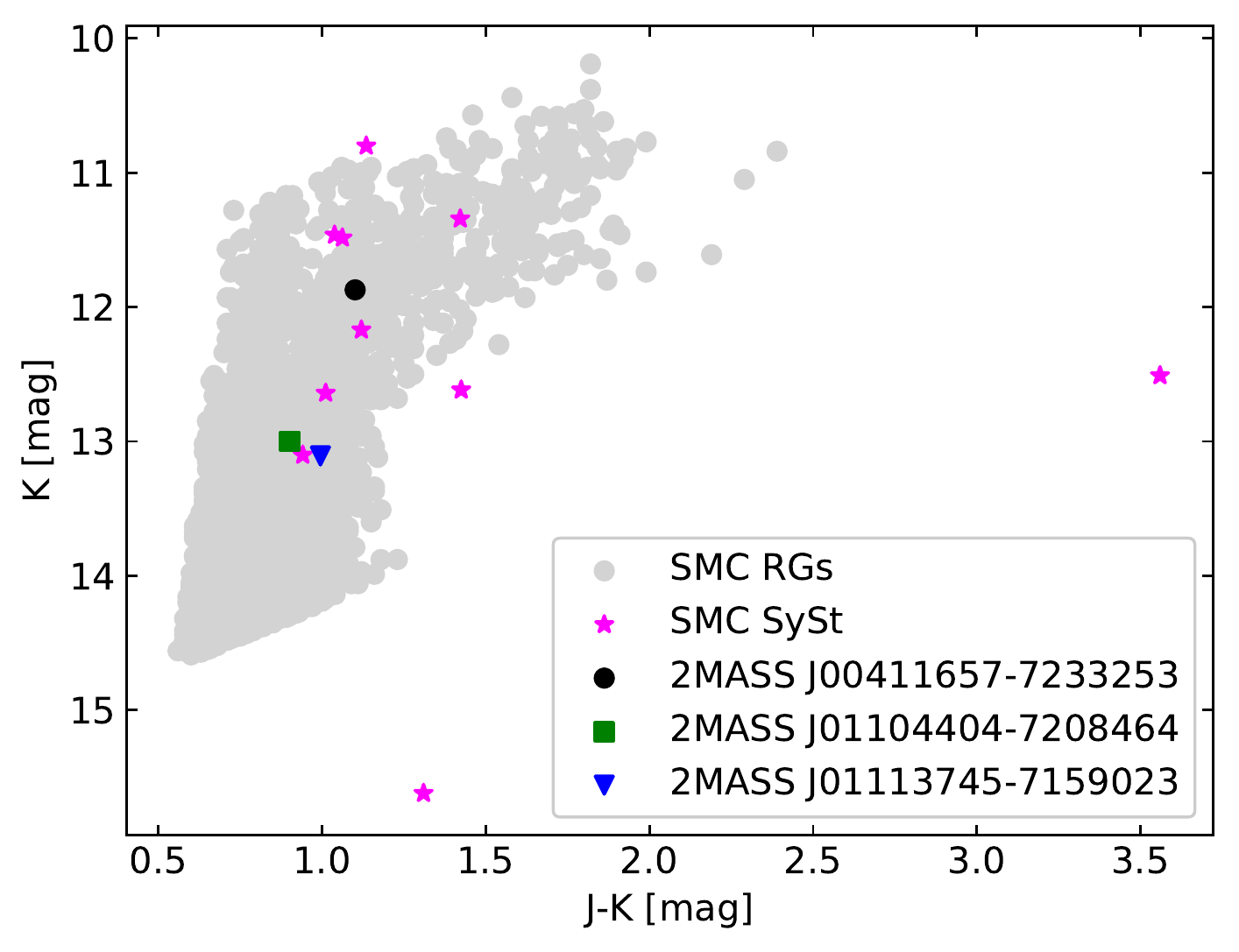}
\caption{Position of the new SySt and known RGs in the SMC on the infrared color-magnitude diagram. The infrared magnitudes are from 2MASS All-Sky Catalog of Point Sources \citep{2000AJ....119.2498J}. The list of known RGs in from \citet{2014MNRAS.442.1680D}.}
\label{CMD}
\end{figure}

Of our sample of ten candidates we found three new SySt: 2MASS J00411657-7233253,	2MASS J01104404-7208464	and	2MASS J01113745-7159023. Each of these new SySt show strong emission lines of the \mbox{H\,{\sc i}} Balmer series, as well as the \mbox{He\,{\sc ii}} 4686 line, in addition to absorption features of K-type giants (Fig.~\ref{fig:salt_spec}). These spectra are very reminiscent of the SALT spectra of the known SMC SySt LIN 9 and LIN 358 (see fig. 3 of \citealt{2014MNRAS.444L..11M}). The SySt natures of 2MASS J01104404-7208464	and 2MASS J01113745-7159023 are further confirmed by the presence of the Raman scattered \mbox{O\,{\sc vi}} 6825 emission line, which is only present in SySt. In the case of 2MASS J00411657-7233253 the emission line ratio $\log$(\mbox{He\,{\sc i}} 6678 / \mbox{He\,{\sc i}} 5876)=0.16 is consistent with a SySt \citep{2017A&A...606A.110I}. In addition, the positions of the three stars in an infrared color-magnitude diagram confirm the presence of a RG in the systems (Fig.~\ref{CMD}).

\begin{figure*}
\centering
\includegraphics[width=0.48\hsize]{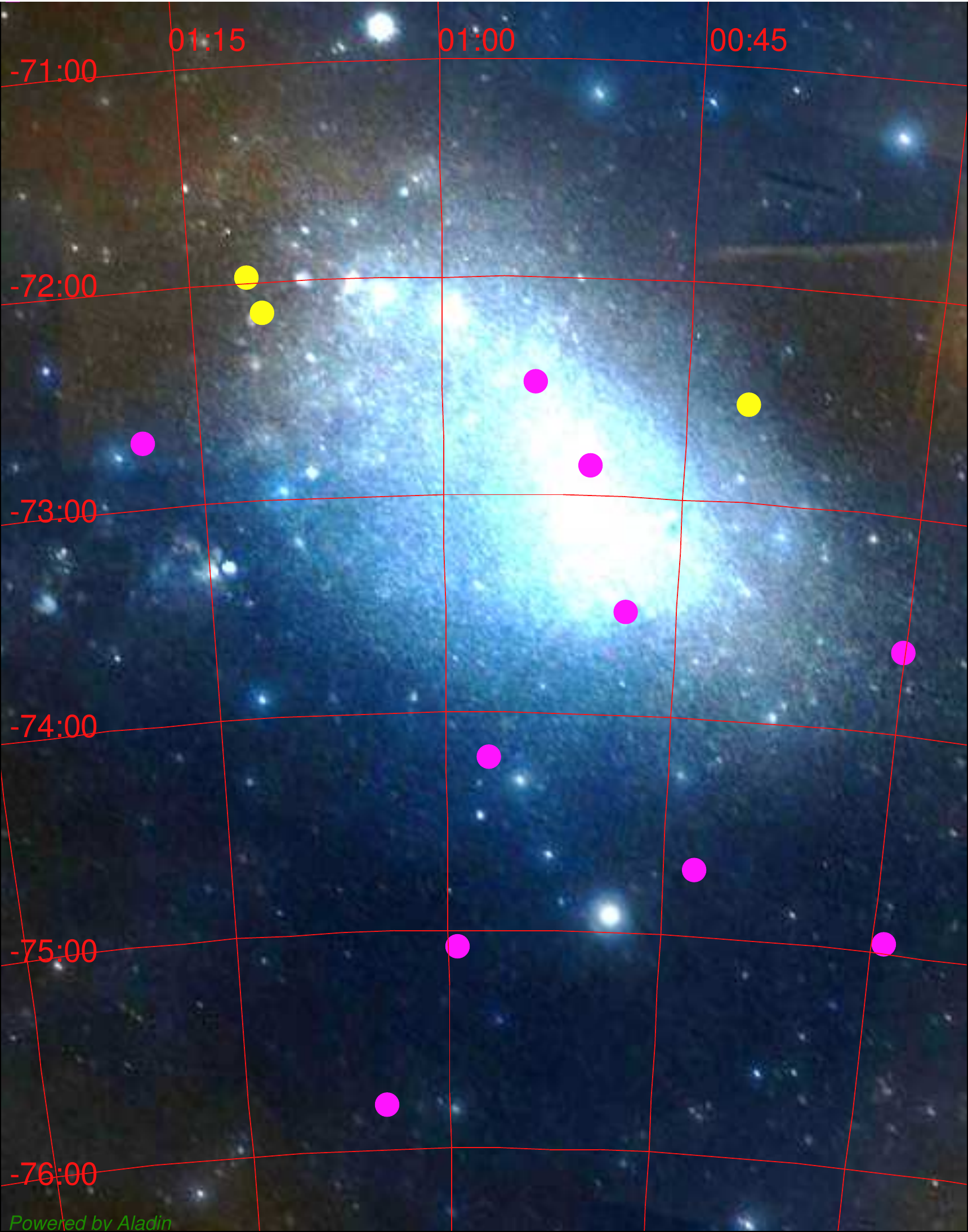}
\includegraphics[width=0.48\hsize]{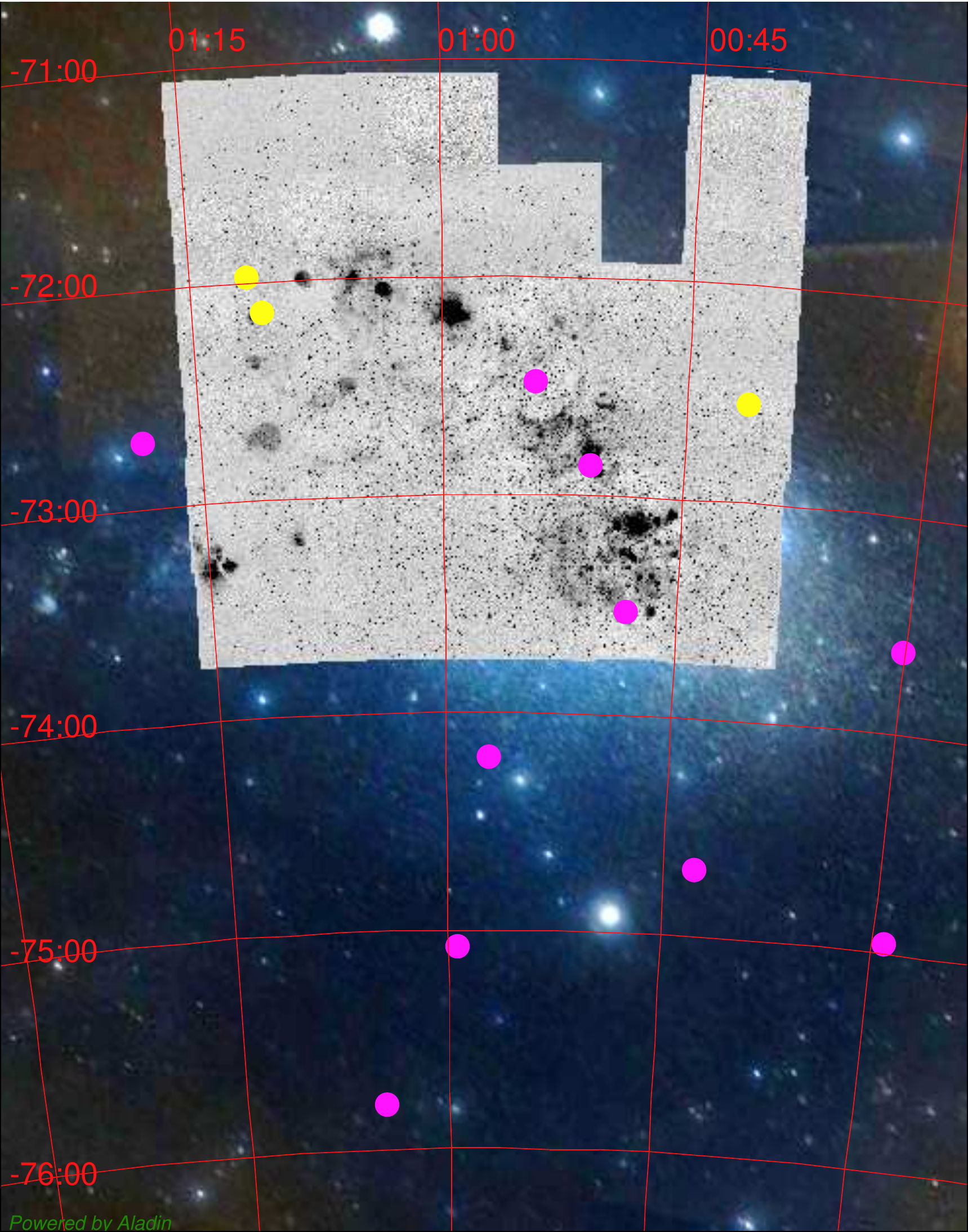}
\caption{Position of the new SySt (yellow points) and ten previously known SySt (magenta circles) in the SMC. Left panel: colored Digitized Sky Survey image of the SMC \citep{1996ASPC..101...88L}. Right panel: Swope H$\alpha$ images plotted over the Digitized Sky Survey image, showing our survey area imaged in 2016 (see text). The list of ten known SySt is from \citet{2014MNRAS.444L..11M}, \citet{2015AcA....65..139H} and references therein. The mosaic of the Swope images was created using Montage toolkit. The images were created using Aladin \citep{2000A&AS..143...33B,2014ASPC..485..277B}. }
\label{positions}
\end{figure*}

The locations of the three new SySt and ten previously known SySt in the SMC are presented in Fig.~\ref{positions}. The black and white inset on the right hand image of Fig.~\ref{positions} shows the survey area we completed in 2016. Almost all of the known SySt are well outside the central regions of the SMC, i.e. where the SMC in the Digitized Sky Survey image is the brightest. This is purely a selection effect, as locating SySt in the crowded central regions is more challenging than in its periphery. The large majority of SySt in the SMC are still not discovered, but as our survey covers the densest and mostly unexplored region of the SMC, we should significantly increase the number of known SySt in the SMC at its conclusion.

\subsection{Variability}

\begin{figure}
\centering
\includegraphics[width=0.95\hsize]{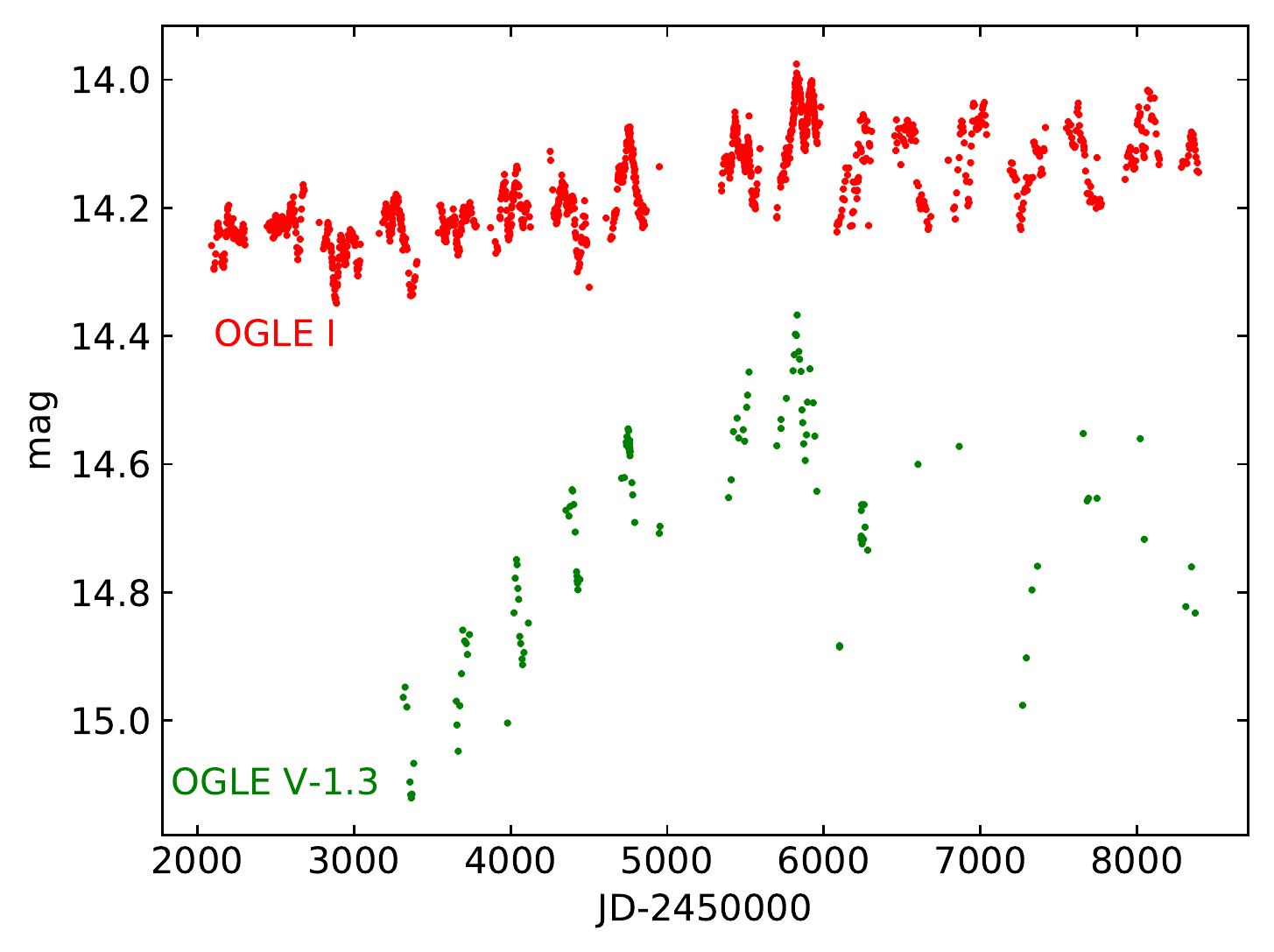}
\caption{OGLE light curve of 2MASS J00411657-7233253.}
\label{ogle_J00411657}
\end{figure}

\begin{figure}
\centering
\includegraphics[width=0.95\hsize]{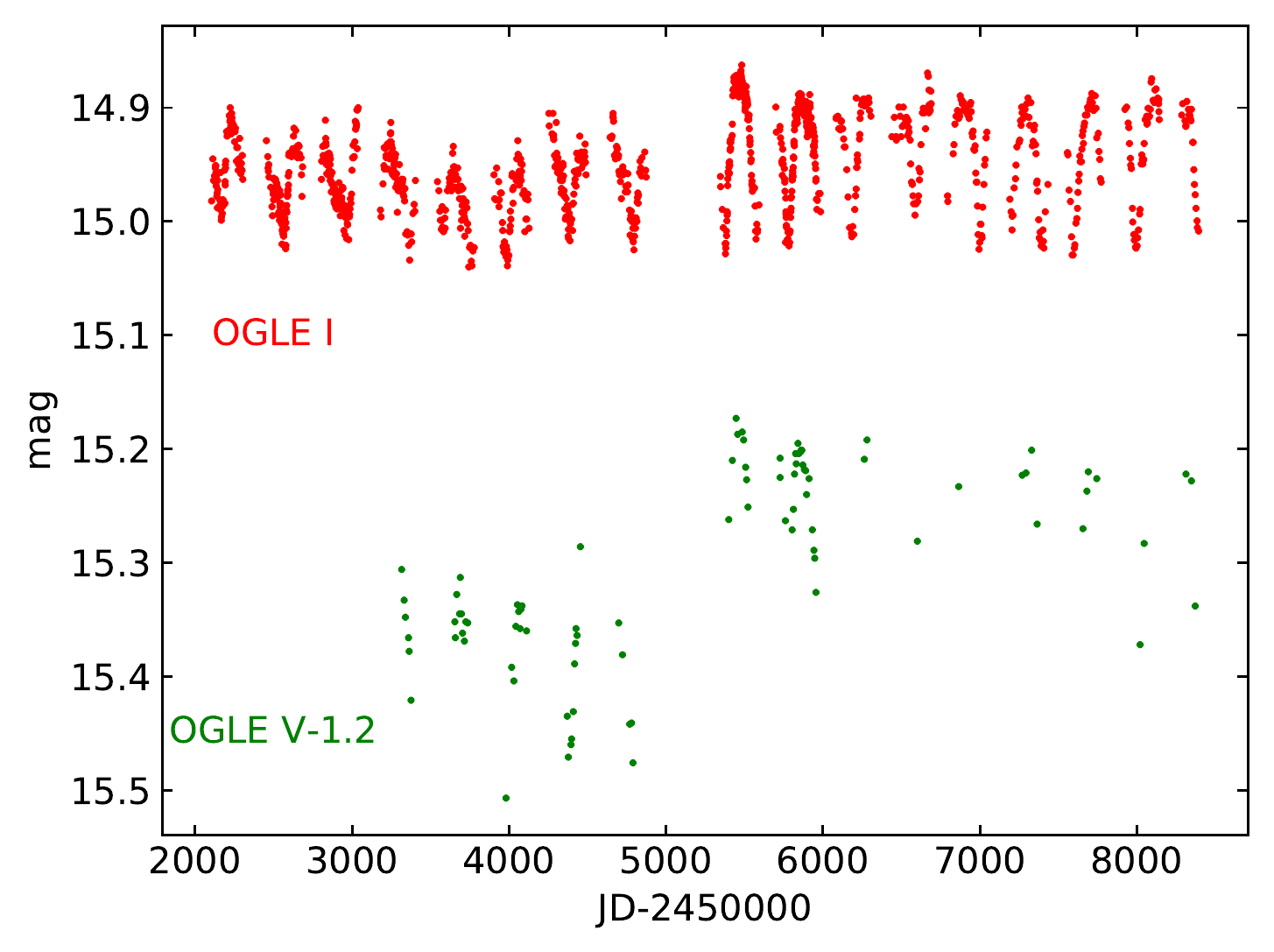}
\caption{OGLE light curve of 2MASS J01104404-7208464.}
\label{ogle_J01104404}
\end{figure}

\begin{figure}
\centering
\includegraphics[width=0.95\hsize]{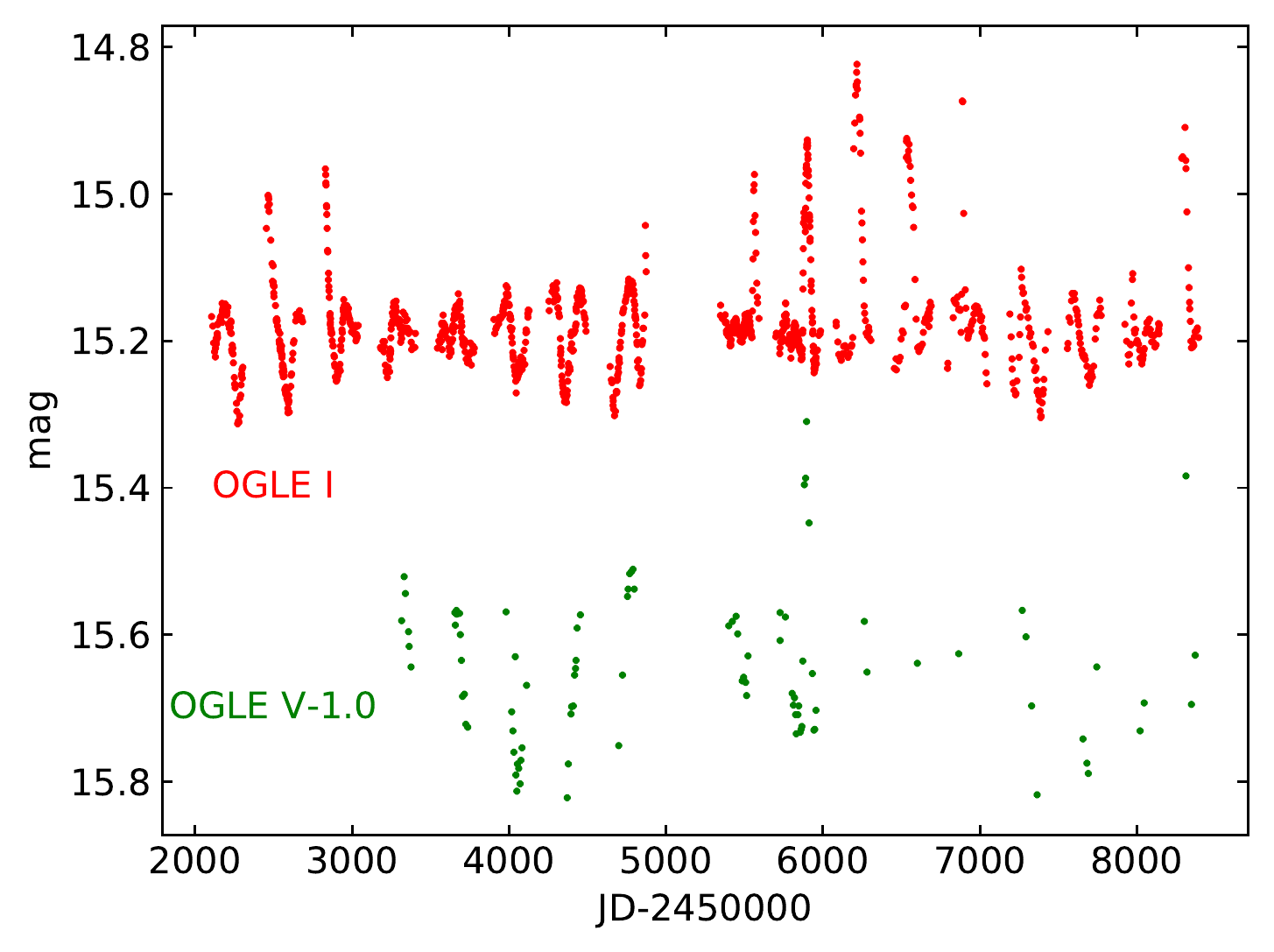}
\caption{OGLE light curve of 2MASS J01113745-7159023.}
\label{ogle_J01113745}
\end{figure}

In order to study the variability of the SySt we collected the Optical Gravitational Lensing Experiment (OGLE) data from \citet{2011AcA....61..217S} catalogue. The OGLE identification numbers of the stars are OGLE-SMC-LPV-03426 for 2MASS J00411657-7233253, OGLE-SMC-LPV-17942 for 2MASS J01104404-7208464 and OGLE-SMC-LPV-18122 for 2MASS J01113745-7159023. The light curves have been updated with the fourth phase of OGLE photometry covering years 2010-2018 \citep{2015AcA....65....1U}. The OGLE light curves are presented in Figs. \ref{ogle_J00411657}-\ref{ogle_J01113745}. We searched for periodicity in the data using a Lomb-Scargle periodogram \citep{1982ApJ...263..835S} and phase dispersion minimization (PDM; \citealt{1978ApJ...224..953S}). The accuracy of periods was estimated by calculating the half-size of a peak in the periodogram. All of the three new SySt were classified previously as OGLE Small Amplitude Red Giants (OSARGs) by \citet{2011AcA....61..217S} based on the data from the third phase of OGLE alone.

The light curve of 2MASS J00411657-7233253 is dominated by a $\sim$70\,d quasi-periodic variability found by  \citet{2011AcA....61..217S}. We also find a $\sim$400\,d period at a low significance and a longer set of data is necessary to confirm this periodicity. If confirmed, the  $\sim$400\,d period could be interpreted as a orbital period. Additionally to the quasi-periodic variability 2MASS J00411657-7233253 showed also an increase in the brightness in $V$ band (Fig.~\ref{ogle_J00411657}). While more data is needed to study the nature of this brightening, it is most probably a result of a small outburst or an active phase.

\begin{figure}[h!]
\centering
\includegraphics[width=0.95\hsize]{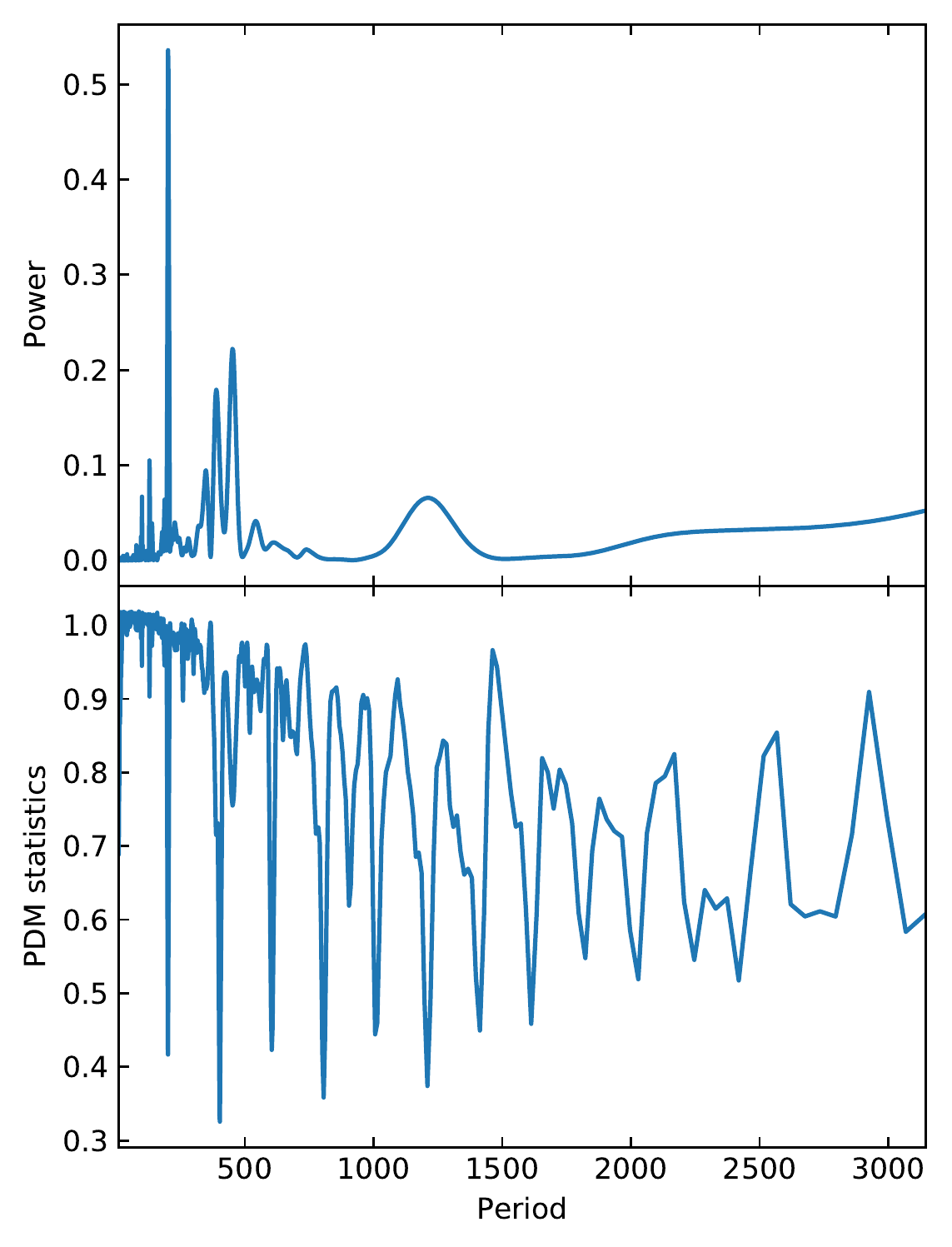}
\caption{Lomb-Scargle periodogram (top) and PDM (bottom) of OGLE I data of 2MASS J01104404-7208464.}
\label{pariodogram}
\end{figure}

\begin{figure}[h!]
\centering
\includegraphics[width=0.95\hsize]{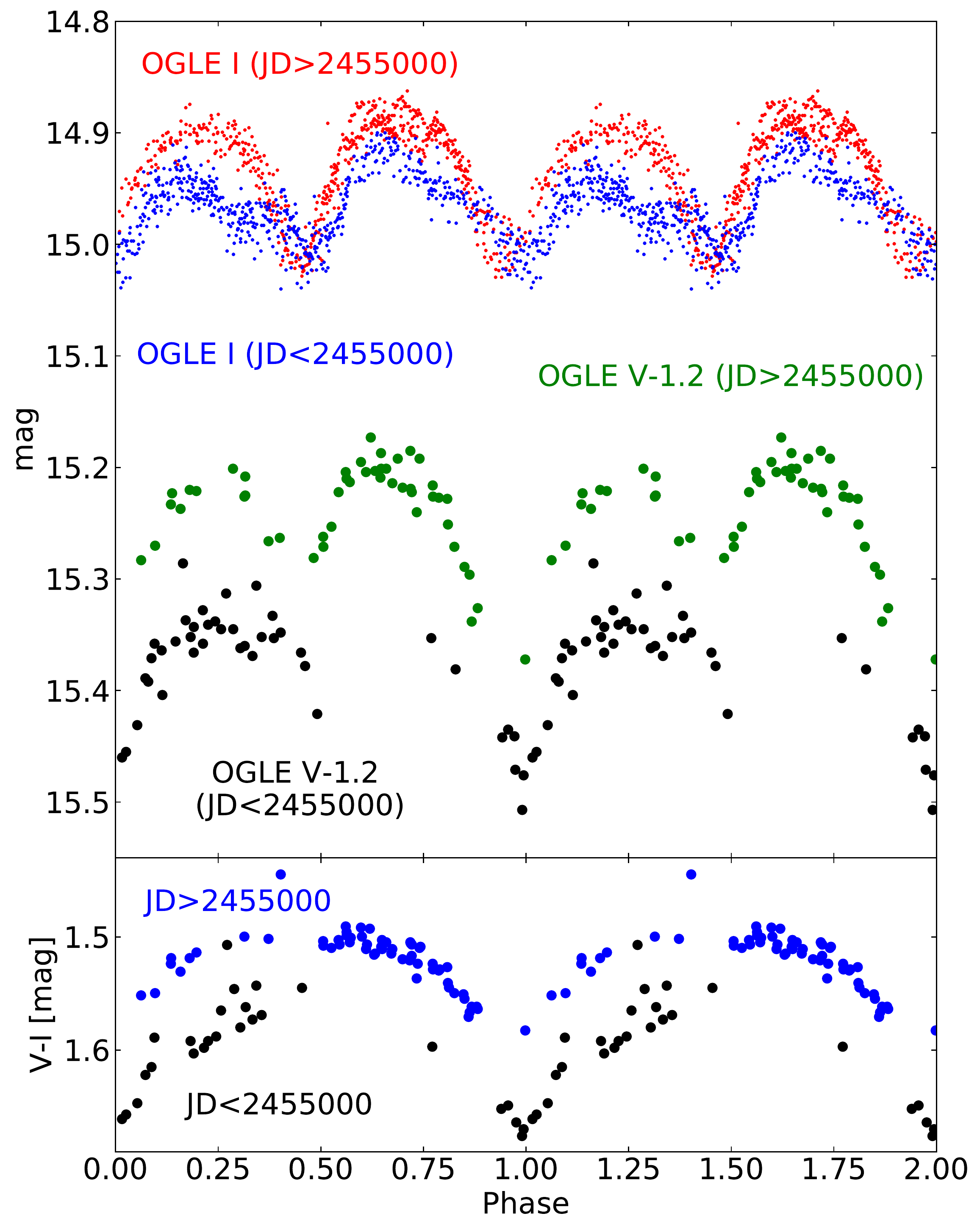}
\caption{Phase plot of OGLE data of 2MASS J01104404-7208464. The data is separated at JD$sim$2455000, at the time of brightening. Only observations obtained on the same date in the two filters are plotted in the case $V-I$ data.}
\label{phaseplot}
\end{figure}

In the case of 2MASS J01104404-7208464 the most prominent period detected in the Lomb-Scargle periodogram is 202\,d, while in the case of PDM the most prominent period is 403\,d (Fig.~\ref{pariodogram}). This suggests that the system shows ellipsoidal variations and the real orbital period is 403\,d, twice as long as the period from the Lomb-Scargle periodogram. The phase plot of 2MASS J01104404-7208464 is presented in Fig.~\ref{phaseplot}. The system showed an increase in the $V$ brightness during a gap in observations at JD$\sim$2455000 (Fig.~\ref{ogle_J01104404}). While this is the time of change from OGLE~III to OGLE~IV and may be associated to changes in the photometric system, the reality of this brightening is confirmed by a change of the shape of $I$ band data in the phaseplot (Fig.~\ref{phaseplot}). The reality of ellipsoidal variations and the 403\,d period is confirmed by the phase plot of $V-I$ color, where the star is getting redder at phase 0 as a result of an eclipse (Fig.~\ref{phaseplot}). The second decrease in brightness at the phase of 0.5 is not associated with change of color, as expected for ellipsoidal variability. Moreover, the RG radius (see Tab.~\ref{tab:properties}) is comparable to its Roche lobe radius (assuming a total mass of $\sim$2 M$_{\sun}$ and P$_\mathrm{orb}$=403\,d). The ephemerides of 2MASS J01104404-7208464 is:

\begin{equation} \label{eq1}
JD_{min}=2454792+(403\pm 8)\times E
\end{equation}

The variability of  2MASS J01113745-7159023 is dominated by a quasi-periodic variability with a period of $\sim$160\,d. However, the maximum of this quasi-periodic variability in some periods of time is higher by a $\sim$0.4~mag in the $I$ band (Fig.~\ref{ogle_J01113745}). This periodic increase in the maximum of variability is reminiscent of variability in a symbiotic X-ray binary GX~1+4 \citep{2017A&A...601A.105I}. In GX~1+4 this behavior is associated with a periastron passage during elliptical orbital motion, i.e. when enhanced mass transfer rate can be expected. If this is the case in 2MASS J01113745-7159023 then the orbital period of the system is $\sim$3500\,d.

\subsection{Stellar components parameters}

In order to estimate WD parameters we estimated the reddening to the SySt by using a map of SMC reddening that includes foreground and average internal SMC reddening (Skowron, D. et al., 2019, in preparation). Because this does not include the SySt intrinsic reddening our estimate might be underestimated. This method gave us A$_V$=0.23~mag in the case of 2MASS J00411657-7233253, A$_V$=0.12~mag in the case of 2MASS J01104404-7208464 and A$_V$=0.11~mag in the case of 2MASS J01113745-7159023. We assumed SMC distance of 62.1\,kpc \citep{2014ApJ...780...59G}.

An effective temperature of WD T$_{\rm WD}$ was estimated using a method of \citet{1981psbs.conf..517I} and emission line fluxes of \mbox{He\,{\sc ii}} 4686, H$_\beta$ and \mbox{He\,{\sc i}} 4471. This resulted in T$_{\rm WD}$=120\,kK for 2MASS J00411657-7233253, T$_{\rm WD}$=160\,kK for 2MASS J01104404-7208464 and T$_{\rm WD}$=230\,kK for 2MASS J01113745-7159023. In the case of 2MASS J01104404-7208464 and  2MASS J01113745-7159023 we assumed that \mbox{He\,{\sc i}} 4471  emission line in negligible since it was not detected in our spectra. 


The luminosity of WD L$_{\rm WD}$ was estimated using H$\beta$ fluxes and eq.~6 from \citet{1997A&A...327..191M}, which gave L$_{\rm WD}$=750\,L$_\odot$ for 2MASS J00411657-7233253, L$_{\rm WD}$=250\,L$_\odot$ for 2MASS J01104404-7208464 and L$_{\rm WD}$=440\,L$_\odot$ for 2MASS J01113745-7159023, respectively. Using the \mbox{He\,{\sc ii}} 4686 fluxes and
eq.~7 from \citet{1997A&A...327..191M} we estimated L$_{\rm WD}$=750\,L$_\odot$ for 2MASS J00411657-7233253, L$_{\rm WD}$=230\,L$_\odot$ for 2MASS J01104404-7208464 and L$_{\rm WD}$=410\,L$_\odot$ for 2MASS J01113745-7159023, respectively. Both of these methods assume  a black body spectrum for the hot compenents, and that both  \mbox{H\,{\sc i}} and \mbox{He\,{\sc ii}} emission lines are produced by photoionization followed by recombination (case B), and they  are accurate to a factor of 2. The temperatures and luminosities of these three white dwarfs are similar to those found for the white dwarfs in other Magellanic SySt.

In order to derive the RG effective temperature we used a relation $T_{\rm RG}=7070/[(J-K)+0.88]$ derived by \citet{1983MNRAS.202...59B}. Using 2MASS All-Sky Catalog of Point Sources \citep{2000AJ....119.2498J} magnitudes we get $T_{\rm RG}$=3570\,K for 2MASS J00411657-7233253, $T_{\rm RG}$=3970\,K for 2MASS J01104404-7208464 and $T_{\rm RG}$=3770\,K for 2MASS J01113745-7159023. The magnitudes used were $J$=12.973~mag and $K$=11.872~mag for 2MASS J00411657-7233253, $J$=13.900~mag	and $K$=12.999~mag for 2MASS J01104404-7208464 and $J$=14.104~mag and $K$=13.109~mag for 2MASS J01113745-7159023. The RG radii were then estimated using the $K$ magnitudes and adopting the bolometric corrections BC$_K$, calculated from  the \mbox{(BC$_K$,($J-K$))} relation given by \citet{1984PASP...96..247B}.

All the stellar parameters derived for the new SySt are presented in Tab.~\ref{tab:properties}.

\begin{table}
\centering
\caption{Properties of stellar components in the new SySt.} \label{tab:properties}
\begin{tabular}{ccccc}
\tablewidth{0pt}
\hline
\hline
2MASS &		T$_{\rm WD}$  &	L$_{\rm WD}$  & $T_{\rm RG}$ & $R_{\rm RG}$ \\
 &		 [kK] &	 [L$_\odot$] & 	[K]  & [R$_\odot$]\\
\hline
 J00411657-7233253 & 120 & 750 & 3570 & 165\\
 J01104404-7208464 & 160 & 230--250 & 3970 & 98\\
 J01113745-7159023 & 230 & 410--440 & 3770 & 93\\
\hline
\end{tabular}
\end{table}

\section{Conclusions}

We presented first reselts of the first search for new SySt using H$\alpha$ and \mbox{He\,{\sc ii}\,4686} narrow-band filters. We found three new SySt in SMC: 2MASS J00411657-7233253,	2MASS J01104404-7208464	and 2MASS J01113745-7159023. This confirms the viability of this method. Two of them, 2MASS J01104404-7208464	and 2MASS J01113745-7159023, showed Raman scattered \mbox{O\,{\sc vi}} 6825 emission line characteristic for SySt. We derived and discussed the physical characteristics for all of them. Their light curves are also presented and analyzed.

2MASS J00411657-7233253 shows quasi-periodic variability and a marginally detectable variability on timescale of $\sim$400\,d. 2MASS J01104404-7208464 is an ellipsoidal variable with an orbital period of 403\,d. 2MASS J01104404-7208464 also experienced a brightening by $\sim$0.2~mag in $V$ band, which was associated with a change of shape of orbitally related variability. 2MASS J01113745-7159023 variability is similar ot the variability observed in GX~1+4, a symbiotic X-ray binary. This suggested an enhanced mass-transfer during a periastron passage during orbital motion. This is particularly interesting because WD in 2MASS J01113745-7159023 is among the hottest in SySt \citep{2010arXiv1011.5657M}, which confirms high  mass transfer rate in the system or a recent outburst.

\acknowledgments

We gratefully acknowledge a generous grant of time on the Swope telescope of the Carnegie Observatories, and the excellent support staff at Las Campanas for all of their help during our observing runs.  KI has been financed by the Polish Ministry of Science and Higher Education Diamond Grant Programme via grant 0136/DIA/2014/43 and by the Foundation for Polish Science (FNP) within the START  program. This study has been partially founded by the  National  Science  Centre,  Poland,  grant OPUS  2017/27/B/ST9/01940. This research is based on observations made with the Southern African Large Telescope (SALT) under programme 2017-2-SCI-040. Polish participation in SALT is funded by grant No. MNiSW DIR/WK/2016/07. The OGLE project has received funding from the National Science Centre, Poland, grant MAESTRO 2014/14/A/ST9/00121 to AU. The Digitized Sky Surveys were produced at the Space Telescope Science Institute under U.S. Government grant NAG W-2166. The images of these surveys are based on photographic data obtained using the Oschin Schmidt Telescope on Palomar Mountain and the UK Schmidt Telescope. The plates were processed into the present compressed digital form with the permission of these institutions. This research made use of Montage. It is funded by the National Science Foundation under Grant Number ACI-1440620, and was previously funded by the National Aeronautics and Space Administration's Earth Science Technology Office, Computation Technologies Project, under Cooperative Agreement Number NCC5-626 between NASA and the California Institute of Technology.

\bibliographystyle{aasjournal} 
\bibliography{references} 

\begin{thebibliography}{}
\expandafter\ifx\csname natexlab\endcsname\relax\def\natexlab#1{#1}\fi
\providecommand{\url}[1]{\href{#1}{#1}}
\providecommand{\dodoi}[1]{doi:~\href{http://doi.org/#1}{\nolinkurl{#1}}}
\providecommand{\doeprint}[1]{\href{http://ascl.net/#1}{\nolinkurl{http://ascl.net/#1}}}
\providecommand{\doarXiv}[1]{\href{https://arxiv.org/abs/#1}{\nolinkurl{https://arxiv.org/abs/#1}}}

\bibitem[{{Belczy{\'n}ski} {et~al.}(2000){Belczy{\'n}ski}, {Miko{\l}ajewska},
  {Munari}, {Ivison}, \& {Friedjung}}]{2000A&AS..146..407B}
{Belczy{\'n}ski}, K., {Miko{\l}ajewska}, J., {Munari}, U., {Ivison}, R.~J., \&
  {Friedjung}, M. 2000, \aaps, 146, 407, \dodoi{10.1051/aas:2000280}

\bibitem[{{Bertin}(2011)}]{2011ASPC..442..435B}
{Bertin}, E. 2011, in Astronomical Society of the Pacific Conference Series,
  Vol. 442, Astronomical Data Analysis Software and Systems XX, ed. I.~N.
  {Evans}, A.~{Accomazzi}, D.~J. {Mink}, \& A.~H. {Rots}, 435

\bibitem[{{Bertin} \& {Arnouts}(1996)}]{1996A&AS..117..393B}
{Bertin}, E., \& {Arnouts}, S. 1996, \aaps, 117, 393,
  \dodoi{10.1051/aas:1996164}

\bibitem[{{Bertin} {et~al.}(2002){Bertin}, {Mellier}, {Radovich}, {Missonnier},
  {Didelon}, \& {Morin}}]{2002ASPC..281..228B}
{Bertin}, E., {Mellier}, Y., {Radovich}, M., {et~al.} 2002, in Astronomical
  Society of the Pacific Conference Series, Vol. 281, Astronomical Data
  Analysis Software and Systems XI, ed. D.~A. {Bohlender}, D.~{Durand}, \&
  T.~H. {Handley}, 228

\bibitem[{{Bessell} \& {Wood}(1984)}]{1984PASP...96..247B}
{Bessell}, M.~S., \& {Wood}, P.~R. 1984, \pasp, 96, 247, \dodoi{10.1086/131328}

\bibitem[{{Bessell} {et~al.}(1983){Bessell}, {Wood}, \&
  {Evans}}]{1983MNRAS.202...59B}
{Bessell}, M.~S., {Wood}, P.~R., \& {Evans}, T.~L. 1983, \mnras, 202, 59,
  \dodoi{10.1093/mnras/202.1.59}

\bibitem[{{Boch} \& {Fernique}(2014)}]{2014ASPC..485..277B}
{Boch}, T., \& {Fernique}, P. 2014, in Astronomical Society of the Pacific
  Conference Series, Vol. 485, Astronomical Data Analysis Software and Systems
  XXIII, ed. N.~{Manset} \& P.~{Forshay}, 277

\bibitem[{{Bonnarel} {et~al.}(2000){Bonnarel}, {Fernique}, {Bienaym{\'e}},
  {Egret}, {Genova}, {Louys}, {Ochsenbein}, {Wenger}, \&
  {Bartlett}}]{2000A&AS..143...33B}
{Bonnarel}, F., {Fernique}, P., {Bienaym{\'e}}, O., {et~al.} 2000, \aaps, 143,
  33, \dodoi{10.1051/aas:2000331}

\bibitem[{{Burgh} {et~al.}(2003){Burgh}, {Nordsieck}, {Kobulnicky}, {Williams},
  {O'Donoghue}, {Smith}, \& {Percival}}]{2003SPIE.4841.1463B}
{Burgh}, E.~B., {Nordsieck}, K.~H., {Kobulnicky}, H.~A., {et~al.} 2003, in
  \procspie, Vol. 4841, Instrument Design and Performance for Optical/Infrared
  Ground-based Telescopes, ed. M.~{Iye} \& A.~F.~M. {Moorwood}, 1463--1471

\bibitem[{{Crawford} {et~al.}(2010){Crawford}, {Still}, {Schellart}, {Balona},
  {Buckley}, {Dugmore}, {Gulbis}, {Kniazev}, {Kotze}, {Loaring}, {Nordsieck},
  {Pickering}, {Potter}, {Romero Colmenero}, {Vaisanen}, {Williams}, \&
  {Zietsman}}]{2010SPIE.7737E..25C}
{Crawford}, S.~M., {Still}, M., {Schellart}, P., {et~al.} 2010, in \procspie,
  Vol. 7737, Observatory Operations: Strategies, Processes, and Systems III,
  773725

\bibitem[{{Dobbie} {et~al.}(2014){Dobbie}, {Cole}, {Subramaniam}, \&
  {Keller}}]{2014MNRAS.442.1680D}
{Dobbie}, P.~D., {Cole}, A.~A., {Subramaniam}, A., \& {Keller}, S. 2014,
  \mnras, 442, 1680, \dodoi{10.1093/mnras/stu926}

\bibitem[{{Gon{\c c}alves} {et~al.}(2008){Gon{\c c}alves}, {Magrini}, {Munari},
  {Corradi}, \& {Costa}}]{2008MNRAS.391L..84G}
{Gon{\c c}alves}, D.~R., {Magrini}, L., {Munari}, U., {Corradi}, R.~L.~M., \&
  {Costa}, R.~D.~D. 2008, \mnras, 391, L84,
  \dodoi{10.1111/j.1745-3933.2008.00561.x}

\bibitem[{{Graczyk} {et~al.}(2014){Graczyk}, {Pietrzy{\'n}ski}, {Thompson},
  {Gieren}, {Pilecki}, {Konorski}, {Udalski}, {Soszy{\'n}ski}, {Villanova},
  {G{\'o}rski}, {Suchomska}, {Karczmarek}, {Kudritzki}, {Bresolin}, \&
  {Gallenne}}]{2014ApJ...780...59G}
{Graczyk}, D., {Pietrzy{\'n}ski}, G., {Thompson}, I.~B., {et~al.} 2014, \apj,
  780, 59, \dodoi{10.1088/0004-637X/780/1/59}

\bibitem[{{Hajduk} {et~al.}(2015){Hajduk}, {Gromadzki}, {Miko{\l}ajewska},
  {Miszalski}, \& {Soszy{\'n}ski}}]{2015AcA....65..139H}
{Hajduk}, M., {Gromadzki}, M., {Miko{\l}ajewska}, J., {Miszalski}, B., \&
  {Soszy{\'n}ski}, I. 2015, \actaa, 65, 139.
\newblock \doarXiv{1508.00089}

\bibitem[{{Iijima}(1981)}]{1981psbs.conf..517I}
{Iijima}, T. 1981, in Photometric and Spectroscopic Binary Systems, ed. E.~B.
  {Carling} \& Z.~{Kopal}, 517

\bibitem[{{I{\l}kiewicz} \& {Miko{\l}ajewska}(2017)}]{2017A&A...606A.110I}
{I{\l}kiewicz}, K., \& {Miko{\l}ajewska}, J. 2017, \aap, 606, A110,
  \dodoi{10.1051/0004-6361/201731497}

\bibitem[{{I{\l}kiewicz} {et~al.}(2018){I{\l}kiewicz}, {Miko{\l}ajewska},
  {Miszalski}, {Koz{\l}owski}, \& {Udalski}}]{2018MNRAS.476.2605I}
{I{\l}kiewicz}, K., {Miko{\l}ajewska}, J., {Miszalski}, B., {Koz{\l}owski}, S.,
  \& {Udalski}, A. 2018, \mnras, 476, 2605, \dodoi{10.1093/mnras/sty365}

\bibitem[{{I{\l}kiewicz} {et~al.}(2017){I{\l}kiewicz}, {Miko{\l}ajewska}, \&
  {Monard}}]{2017A&A...601A.105I}
{I{\l}kiewicz}, K., {Miko{\l}ajewska}, J., \& {Monard}, B. 2017, \aap, 601,
  A105, \dodoi{10.1051/0004-6361/201630021}

\bibitem[{{Jarrett} {et~al.}(2000){Jarrett}, {Chester}, {Cutri}, {Schneider},
  {Skrutskie}, \& {Huchra}}]{2000AJ....119.2498J}
{Jarrett}, T.~H., {Chester}, T., {Cutri}, R., {et~al.} 2000, \aj, 119, 2498,
  \dodoi{10.1086/301330}

\bibitem[{{Kniazev} {et~al.}(2009){Kniazev}, {V{\"a}is{\"a}nen}, {Whitelock},
  {Menzies}, {Feast}, {Grebel}, {Buckley}, {Hashimoto}, {Loaring},
  {Romero-Colmenero}, {Sefako}, {Burgh}, \& {Nordsieck}}]{2009MNRAS.395.1121K}
{Kniazev}, A.~Y., {V{\"a}is{\"a}nen}, P., {Whitelock}, P.~A., {et~al.} 2009,
  \mnras, 395, 1121, \dodoi{10.1111/j.1365-2966.2009.14617.x}

\bibitem[{{Kobulnicky} {et~al.}(2003){Kobulnicky}, {Nordsieck}, {Burgh},
  {Smith}, {Percival}, {Williams}, \& {O'Donoghue}}]{2003SPIE.4841.1634K}
{Kobulnicky}, H.~A., {Nordsieck}, K.~H., {Burgh}, E.~B., {et~al.} 2003, in
  \procspie, Vol. 4841, Instrument Design and Performance for Optical/Infrared
  Ground-based Telescopes, ed. M.~{Iye} \& A.~F.~M. {Moorwood}, 1634--1644

\bibitem[{{Lang} {et~al.}(2010){Lang}, {Hogg}, {Mierle}, {Blanton}, \&
  {Roweis}}]{2010AJ....139.1782L}
{Lang}, D., {Hogg}, D.~W., {Mierle}, K., {Blanton}, M., \& {Roweis}, S. 2010,
  \aj, 139, 1782, \dodoi{10.1088/0004-6256/139/5/1782}

\bibitem[{{Lasker} {et~al.}(1996){Lasker}, {Doggett}, {McLean}, {Sturch},
  {Djorgovski}, {de Carvalho}, \& {Reid}}]{1996ASPC..101...88L}
{Lasker}, B.~M., {Doggett}, J., {McLean}, B., {et~al.} 1996, in Astronomical
  Society of the Pacific Conference Series, Vol. 101, Astronomical Data
  Analysis Software and Systems V, ed. G.~H. {Jacoby} \& J.~{Barnes}, 88

\bibitem[{{Mikolajewska}(2010)}]{2010arXiv1011.5657M}
{Mikolajewska}, J. 2010, ArXiv e-prints.
\newblock \doarXiv{1011.5657}

\bibitem[{{Miko{\l}ajewska}(2012)}]{2012BaltA..21....5M}
{Miko{\l}ajewska}, J. 2012, Baltic Astronomy, 21, 5,
  \dodoi{10.1515/astro-2017-0352}

\bibitem[{{Mikolajewska} {et~al.}(1997){Mikolajewska}, {Acker}, \&
  {Stenholm}}]{1997A&A...327..191M}
{Mikolajewska}, J., {Acker}, A., \& {Stenholm}, B. 1997, \aap, 327, 191

\bibitem[{{Miko{\l}ajewska} {et~al.}(2014){Miko{\l}ajewska}, {Caldwell}, \&
  {Shara}}]{2014MNRAS.444..586M}
{Miko{\l}ajewska}, J., {Caldwell}, N., \& {Shara}, M.~M. 2014, \mnras, 444,
  586, \dodoi{10.1093/mnras/stu1480}

\bibitem[{{Miko{\l}ajewska} {et~al.}(2015){Miko{\l}ajewska}, {Shara},
  {Caldwell}, {Drozd}, {I{\l}kiewicz}, \& {Zurek}}]{2015EAS....71..199M}
{Miko{\l}ajewska}, J., {Shara}, M.~M., {Caldwell}, N., {et~al.} 2015, in EAS
  Publications Series, Vol.~71, EAS Publications Series, 199--204

\bibitem[{{Miko{\l}ajewska} {et~al.}(2017){Miko{\l}ajewska}, {Shara},
  {Caldwell}, {I{\l}kiewicz}, \& {Zurek}}]{2017MNRAS.465.1699M}
{Miko{\l}ajewska}, J., {Shara}, M.~M., {Caldwell}, N., {I{\l}kiewicz}, K., \&
  {Zurek}, D. 2017, \mnras, 465, 1699, \dodoi{10.1093/mnras/stw2937}

\bibitem[{{Miszalski} \& {Miko{\l}ajewska}(2014)}]{2014MNRAS.440.1410M}
{Miszalski}, B., \& {Miko{\l}ajewska}, J. 2014, \mnras, 440, 1410,
  \dodoi{10.1093/mnras/stu292}

\bibitem[{{Miszalski} {et~al.}(2013){Miszalski}, {Miko{\l}ajewska}, \&
  {Udalski}}]{2013MNRAS.432.3186M}
{Miszalski}, B., {Miko{\l}ajewska}, J., \& {Udalski}, A. 2013, \mnras, 432,
  3186, \dodoi{10.1093/mnras/stt673}

\bibitem[{{Miszalski} {et~al.}(2014){Miszalski}, {Miko{\l}ajewska}, \&
  {Udalski}}]{2014MNRAS.444L..11M}
---. 2014, \mnras, 444, L11, \dodoi{10.1093/mnrasl/slu098}

\bibitem[{{Munari} \& {Zwitter}(2002)}]{2002A&A...383..188M}
{Munari}, U., \& {Zwitter}, T. 2002, \aap, 383, 188,
  \dodoi{10.1051/0004-6361:20011724}

\bibitem[{{O'Donoghue} {et~al.}(2006){O'Donoghue}, {Buckley}, {Balona},
  {Bester}, {Botha}, {Brink}, {Carter}, {Charles}, {Christians}, {Ebrahim},
  {Emmerich}, {Esterhuyse}, {Evans}, {Fourie}, {Fourie}, {Gajjar}, {Gordon},
  {Gumede}, {de Kock}, {Koeslag}, {Koorts}, {Kriel}, {Marang}, {Meiring},
  {Menzies}, {Menzies}, {Metcalfe}, {Meyer}, {Nel}, {O'Connor}, {Osman}, {Du
  Plessis}, {Rall}, {Riddick}, {Romero-Colmenero}, {Potter}, {Sass},
  {Schalekamp}, {Sessions}, {Siyengo}, {Sopela}, {Steyn}, {Stoffels},
  {Scholtz}, {Swart}, {Swat}, {Swiegers}, {Tiheli}, {Vaisanen}, {Whittaker}, \&
  {van Wyk}}]{2006MNRAS.372..151O}
{O'Donoghue}, D., {Buckley}, D.~A.~H., {Balona}, L.~A., {et~al.} 2006, \mnras,
  372, 151, \dodoi{10.1111/j.1365-2966.2006.10834.x}

\bibitem[{{Rodr{\'{\i}}guez-Flores} {et~al.}(2014){Rodr{\'{\i}}guez-Flores},
  {Corradi}, {Mampaso}, {Garc{\'{\i}}a-Alvarez}, {Munari}, {Greimel},
  {Rubio-D{\'{\i}}ez}, \& {Santander-Garc{\'{\i}}a}}]{2014A&A...567A..49R}
{Rodr{\'{\i}}guez-Flores}, E.~R., {Corradi}, R.~L.~M., {Mampaso}, A., {et~al.}
  2014, \aap, 567, A49, \dodoi{10.1051/0004-6361/201323182}

\bibitem[{{Scargle}(1982)}]{1982ApJ...263..835S}
{Scargle}, J.~D. 1982, \apj, 263, 835, \dodoi{10.1086/160554}

\bibitem[{{Soszy{\'n}ski} {et~al.}(2011){Soszy{\'n}ski}, {Udalski},
  {Szyma{\'n}ski}, {Kubiak}, {Pietrzy{\'n}ski}, {Wyrzykowski}, {Ulaczyk},
  {Poleski}, {Koz{\l}owski}, \& {Pietrukowicz}}]{2011AcA....61..217S}
{Soszy{\'n}ski}, I., {Udalski}, A., {Szyma{\'n}ski}, M.~K., {et~al.} 2011,
  \actaa, 61, 217.
\newblock \doarXiv{1109.1143}

\bibitem[{{Stellingwerf}(1978)}]{1978ApJ...224..953S}
{Stellingwerf}, R.~F. 1978, \apj, 224, 953, \dodoi{10.1086/156444}

\bibitem[{{Udalski} {et~al.}(2015){Udalski}, {Szyma{\'n}ski}, \&
  {Szyma{\'n}ski}}]{2015AcA....65....1U}
{Udalski}, A., {Szyma{\'n}ski}, M.~K., \& {Szyma{\'n}ski}, G. 2015, \actaa, 65,
  1.
\newblock \doarXiv{1504.05966}

\end{thebibliography}

\end{document}